\documentclass[pdflatex,sn-mathphys-num,iicol]{sn-jnl}

\usepackage{graphicx}%
\usepackage{multirow}%
\usepackage{amsmath,amssymb,amsfonts}%
\usepackage{amsthm}%
\usepackage{mathrsfs}%
\usepackage[title]{appendix}%
\usepackage{xcolor}%
\usepackage{textcomp}%
\usepackage{manyfoot}%
\usepackage{booktabs}%
\usepackage{algorithm}%
\usepackage{algorithmicx}%
\usepackage{algpseudocode}%
\usepackage{listings}%
\usepackage{amsmath}
\usepackage{verbatim}

\theoremstyle{thmstyleone}%
\theoremstyle{thmstyletwo}%

\theoremstyle{thmstylethree}%

\begin{document}

\title[Article Title]{Modeling the effectiveness of radiation shielding materials for astronaut protection on Mars}

\author[1,2*]{Dionysios Gakis}
\author[3,2]{Dimitra Atri}
\affil[1]{Department of Physics, University of Patras, Patras, Rio, 26504, Greece}
\affil[2]{Blue Marble Space Institute of Science, Seattle, WA, 98104, USA}
\affil[3]{Center for Astrophysics and Space Science, New York University Abu Dhabi, PO Box 129188, Abu Dhabi, UAE}
\affil[*]{dgakis@upnet.gr}








\abstract{The surface of Mars is bombarded by energetic charged particles of solar and cosmic origin with little shielding compared to Earth. As space agencies are planning for crewed missions to the red planet, a major concern is the impact of ionizing radiation on astronaut health. Keeping exposure below acceptable radiation dose levels is crucial for the health of the crew. In this study, our goal is to understand the radiation environment of Mars and describe the main strategies to be adopted to protect astronauts from the harmful impacts of cosmic radiation. Specifically, we investigate the shielding properties of various materials in the Martian radiation field using the Geant4 numerical model, after validating its accuracy with in-situ instrument measurements by MSL RAD. Our results indicate that composite materials such as types of plastic, rubber or synthetic fibers, have a similar response against cosmic rays and are the best shields. Martian regolith has an intermediate behavior and therefore could be used as an additional practical option. We show that the most widely used aluminum could be helpful when combined with other low atomic number materials. 
}




\maketitle

\section{Introduction}\label{sec1}

Ionizing radiation is one of the major hazards astronauts (referring to human crew, also known as cosmonauts, taikonauts etc.) must face in long-duration space missions. Other dangers include microgravity, absence of oxygen, debris and isolation, but ionizing radiation ranks as one of the most threatening because of its ability to penetrate shielding and cause biological damage \cite{atri2014cosmic}. For this reason, awareness of space radiation and its harmful effects, is widely spread among the scientific community and is a highly interdisciplinary endeavour.

Unlike the Earth, Mars lacks a thick atmosphere and a magnetic field able to deflect particles and reduce radiation doses to low levels. Hence, we must look for ways to protect human crews from these hazardous charged particles. While radiation dose rate on the surface of Mars is of course much lower than when in deep space, the already accumulated doses in astronauts' bodies during their transfer to and from Mars make it inevitable to search for strategies of mitigating exposure on Mars too.

Several techniques have been developed to protect astronauts from absorbing immense amounts of radiation (e.g. \cite{Wilson:1997}), the main being passive shielding, which takes advantage of the energy loss through the mass of a shielding material. In literature so far, aluminum and polyethylene have been well tested in the Martian radiation field \cite{Durante:2011,Slaba:2013}. Besides, the in-situ regolith has also been suggested due to practicability \cite{Simonsen:1991,Kim:1998}. Other works like \cite{Naito:2020} simulate radiation doses behind other materials too, but they are mostly limited to ISS or the deep space environment.

The purpose of this study is to outline the suggested radiation protection techniques, focusing on testing the effectiveness of various materials against radiation, specifically on the surface of Mars. We re-examine the effectiveness of materials commonly used in space missions and try novel ones. Additionally, we compare the efficiency of combinations of shielding materials, trying to benefit from each one's unique properties. To that end, we employed the computational program Geant4 to simulate the propagation of energetic charged particles through matter.

This paper is organized as follows. Section 2 outlines the radiation environment on the surface of Mars and Section 3 consists of the main shielding strategies agencies have adopted for radiation mitigation. We then describe our computational method in Section 4, present our results in Section 5 and discuss their validity in Section 6. Finally, we provide summary and conclusions in the last section (Section 7).

\section{Radiation environment}

Astrophysical radiation consists of photons and charged particles, which can further be categorized as ionizing and non-ionizing. Both kinds of radiation can damage living and non-living systems. Non-ionizng radiation includes for instance secondary neutrons that can permeate any kind of thin shield, and thus are difficult to shield against. Ionizing radiation passes through substances and alters them, producing a cascade of secondary particles, increasing with increasing energy of incident radiation. For example, a single proton may produce through inelastic interactions with the matter secondary particles such as kaons, pions, muons and other unstable particles, eventually resulting in the exposure to neutrons, electrons, muons and other particles, which are known to cause biological damage and pose health risks to astronauts. Therefore, one of the main concerns during space missions is protecting astronauts from the devastating effects of ionizing radiation.

\subsection{Radiation sources}
Radiation in interplanetary space and on the surface of Mars originates from the Sun and from Galactic sources. Energetic processes in the solar corona release energetic charged particles called Solar Energetic Particles (SEPs) and background flux of charged particles from Galactic sources known as Galactic Cosmic Rays (GCR). There may be local sources of radiation in form of radioactivity in the Martian regolith, but their dose is much smaller and extraterrestrial ones are the main source of health hazard; consequently we focus on the latter.

SEPs are associated with solar flares and coronal mass ejections (CMEs) \cite{Reames:1999}. They consist mainly of protons (90-95\%), followed by helium nuclei (5-8\%), along with some heavier nuclei (1\%). Clearly, these percentages can only be rough, as exact compositions differ a lot between events or even during a single event \cite{Tylka:1999}. The typical energy spectra of SEP events cover the range of some keV to hundreds of MeV, softer than GCR. In some extreme cases, they can go up to 10 GeV \cite{tylka2009new, atri2020stellar}. 

Solar Particle Events (SPEs) are rare and generally occur in phase with the solar cycle, at times of maximum solar activity, despite the fact that some of them occur at any time of the solar activity. They are produced during solar eruptions, at which particles are accelerated to relativistic energies by the plasma \cite{Kallenrode:2003}. Nonetheless, the exact mechanism of SPEs' creation and the following effects in the space weather have not yet been definitely determined. In other words, they are quite unpredictable and might differ significantly by duration, strength and fluence. Historically recorded events of exceptionally intense solar activity include the Carrington event (September 1859), along with the events of February 1956, August 1972 and September 1989. It appears that they used to happen more often in the past; caution is then needed when extrapolating to predict future events.

The other type of astrophysical radiation, GCR, has galactic origins. GCR are accelerated by supernova shocks and their trajectories are bent by interstellar magnetic fields before entering the solar system. As a result, they are considered to have isotropic flux (i.e. coming from every direction in the sky), and be nearly stable through thousands of years. Their composition is similar, but different from SEPs': 87\% protons, 12\% alpha particles, 1\% heavy nuclei ($Z>2$) \cite{Simpson:1983}. Some electrons are also observed (1\%), but are often ignored due to their low abundance in contrast to solar electrons. GCR have an average energy of some GeV per nucleon, but it is possible that galactic particles are accelerated at kinetic energies as high as some PeV \cite{Blasi:2013}.

The so-called HZE (high charge and energy) ions, coming from both radiation sources, contribute significantly to radiation doses. HZE are in fact atoms of medium atomic numbers (up until Fe), stripped off their electrons, travelling freely through space at speeds near the speed of light. Despite their small abundances, they are characterized by strong ionizing and penetration power. They have the ability to penetrate deep inside matter, and sometimes break apart inside materials, thus creating a cascade of secondary, also highly energetic and charged, particles, dangerous for human health \cite{Cucinotta:2014}. Shielding from this kind of cosmic particles is an extremely difficult task.

SPEs and GCR are anti-correlated at short timescales (months to years). At periods of minimum flux of SPEs, GCR appear to have a peak in their intensity, and vice versa (e.g. \cite{Cane:1999}). The proposed explanation to this behavior is the following: at periods of minimum solar activity, the intensity of interplanetary magnetic field (IMF) is reduced. As a result, it does not block efficiently the incoming galactic cosmic radiation, which penetrates well inside the solar region. The Badhwar-O’Neil 2010 model \cite{Neill:2010} is widely used to account for the energy loss of GCR as they travel through the heliosphere, considering factors like deceleration, diffusion and convection. In this model, a parameter $\Phi$ of units MV is defined, in a way that a small value of $\Phi$ corresponds to a large GCR flux. Typical values for solar minimum condition are $\Phi=400-450$ MV and $\Phi=1000$ MV for a solar maximum, or even $\Phi=1800$ MV for a very strong solar event \cite{Neill:2006}. 

\subsection{Conditions on Mars}
There have not been enough comprehensive measurements of the radiation environment of Mars from orbit and from the surface. The Mars Science Laboratory's (MSL) Radiation Assessment Detector (RAD) instrument \cite{Hassler:2012}, launched as part of Curiosity Rover, is currently measuring the radiation levels on the surface of Mars. Located inside the Gale Crater (about 4 km deeper from the mean Martian surface - elevation is estimated by Mars Orbiter Laser Altimeter, \cite{Smith:1999}), it measures since August 2012 charged and neutral particles by GCR and SPEs, accounting for both downward and albedo particles \cite{Ehresmann:2014,Kohler:2014,Hassler:2014}. It is a highly valuable experiment, since it provides results for dose rates, zenith angle dependencies, diurnal and seasonal variations, among others \cite{Rafkin:2014,Guo:2015,Wimmer:2015}. An extensive review on the knowledge acquired by RAD is given in \cite{Guo:2021}.

Observations of Martian environment are available from satellites and surface detectors (e.g. the Marie instrument on board 2001 Mars Odyssey \cite{Atwell:2004}), but they are limited in time (last decades). As for in-situ data, they are only available during the last few years with MSL RAD. Because of that, our knowledge about the situation on Mars is limited, which affects the total planning of missions. We then give a short review of the actual known conditions on Mars, coming from long-term observations or models, and as validated by RAD measurements.

As a fist approach, the bulk of Mars reduces radiation exposure to half for an astronaut placed on Mars mission in comparison to when located in deep space. All upward-directed rays are absorbed by the mass of the planet. Hence, radiation arrives isotropically over a solid angle of $2\pi$ steradians. 

Secondly, Mars is surrounded by a very thin atmosphere, composed almost entirely by carbon dioxide (CO$_2$) with traces of nitrogen, water vapor, and oxygen. At a mean column density of 20 g cm$^{-2}$ (Gale Crater), it is two orders of magnitude thinner than Earth's atmosphere \cite{Williams:2004}. This very thin atmosphere offers some protection from radiation though, preventing particles of low energy to reach the surface. The proton energy cutoff has been found to be around 150-160 MeV (it may vary between 155 to 178 MeV depending on the column density, \cite{Guo:2019}). In other words, only particles with energies above this threshold may pass through the Martian atmosphere. Some SPEs have these low energies and are thus retained away from the surface due to ionization processes, as opposed to GCR, which have generally more energetic spectra. GCR interact with the molecules of atmosphere before being stopped, and they create secondary particles which can traverse all of the atmospheric layers.

The direction of the incoming solar particles plays a role as well. Atmosphere's protection increases at low altitudes. Near the horizon, particles have to travel through longer paths within the atmosphere when compared to radiation coming from near the zenith. To give a quantitative view of the importance of the particle's origin, the density of the atmosphere at zero elevation for zenith angle $z=0$ is 16.0 g cm$^{-2}$, whereas it increases to 120 g cm$^{-2}$ at $z=90^{\circ}$ \cite{Townsend:2010}, so the additional protection is obvious. The above hold true for SPEs at most. GCR flux is almost kept intact over varying incoming angles, because of their relatively higher energies. The elevation has a similar effect, as corresponding radiation paths are shorter for example at the top of a mountain, compared to the mean elevation level or deep inside a crater. It becomes clear that radiation varies from location to location \cite{Saganti:2004}. It is then crucial to determine precisely the area of the human activity during the astronaut stay, as we further discuss in Section 3.2.

Other phenomena that occur inside the atmosphere of Mars have a minor effect in the radiation condition on the planet. For example, \cite{Guo:2017} compared the short-term modulations of GCR-induced doses caused by local atmospheric pressure changes with the long-term solar modulation of GCR, and concluded that although the high-pressure atmosphere results in lower radiation levels, surface doses are mainly influenced by the second effect. Similarly, dust storms have insignificant effect in radiation doses, because dust density is much lower than atmospheric mean density \cite{Keating:2006, Norman:2014}. The effect by diurnal thermal tides and seasonal atmospheric variations is also statistically smaller than the heliospheric modulation of GCR \cite{Rafkin:2014, Guo:2015}.

At last, Mars lacks a global magnetic field \cite{Acuna:1998}. There is not an intrinsic magnetic field like Earth's. Only various regional magnetic fields have been detected, with intensities of 30-60 nT. These magnetic fields are remnants of the planet's ancient magnetosphere. In Earth, the analog situation is an active dynamo, located in the core of planet, which covers its bulk. As a result, charged particles easily impinge on the top layers of the Martian atmosphere, since there is not a mechanism (like the magnetic lines) able to deflect them and divert them to trajectories away from the surface.

The principal radiation source of radiation on the surface of Mars is GCR. GCR are always present and appear as an isotropic, highly predictable and nearly constant background radiation, hard to shield against \cite{Hassler:2014}. Although their overall daily flux may be low, they become dangerous when they accumulate during long missions. SPEs are less frequent, less energetic and can easily be attenuated by the atmosphere, but can potentially deliver great doses in a short period of time and cause illness or even death \cite{Townsend:2011}. 

Summarizing, the radiation field on the red planet comprises mainly of primary GCR, along with secondaries generated in atmosphere and scattered from surface \cite{Saganti:2002}. Furthermore, the radiation environment on Mars is similar to the one in the vicinity of ISS, as it was evaluated by RAD. Despite the fact that charged particles trapped by the Van Allen belts shape the radiation environment around Earth, which is not the case on Mars, their mean quality factors (see Appendix) are similar. The total dose rate is around 0.240 mGy/day for ISS region and 0.213 mGy/day on Mars \cite{Matthia:2017}.

\section{Shielding against Space Radiation}
Outside of the protective shield of the earth’s atmosphere and magnetosphere, astronauts are exposed to enhanced radiation dose, which is also the case for the Martian surface where the atmosphere is thin and there is no global magnetic field. This exposure might be acute, meaning that substantial amounts of radiation strike during a short time interval, or chronic (low-level radiation doses, which become dangerous when they accumulate in large numbers). We would place SPEs in the first kind of exposure, and GCR in the second one. Obviously, both should be addressed, as they can lead to devastating effects to the astronauts' health (e.g. \cite{Kennedy:2014}). The sudden particle flux by a major solar event might be lethal \cite{Wu:2009}, and the constant hit of GCR particles on human cells perhaps leads into long-term sicknesses, like cancer \cite{Cucinotta:2006}. Given the above, it comes as no surprise that scientists and space agencies have developed many strategies against radiation. 

The radiation levels on the surface of Mars, at mean quality factor (see Appendix) $<Q>\, =2-3$ (e.g.  \cite{Hassler:2014}, \cite{Matthia:2017}, \cite{Zeitlin:2019}), are typically within acceptable limits \cite{national2021space} (still, a solar event may rise absorbed doses to critical levels). Nevertheless, a typical journey to Mars takes 2 to 3 years. During that time period of space travel in the interplanetary region, astronauts will be exposed to substantial doses of radiation. These already accumulated doses make it necessary to try to mitigate radiation levels on Mars as well, through some strategies, in order to maintain exposure levels below allowed career limits \cite{ICRP:123}, \cite{national2021space}. We present strategies in the following parts which are crucial to reducing radiation exposure with the goal of maintaining radiation levels below the career exposure limit.

\subsection{Shielding techniques}
At present, two main radiation shielding techniques on space have been proposed and implemented, to name passive shielding and active shielding. Passive shielding utilizes a bulk material, inside of which the majority of particles' energy is deposited before reaching the human bodies. Instead, the main idea behind active shielding is to use electrostatic or plasma shields and confined or unconfined magnetic fields to avert charged particles from their original direction, thus minimizing or avoiding irradiation \cite{Spillantini:2008}. Active shielding is a practical solution, considering the weight launch limitations of passive shielding. However, it is still in quite theoretical stages and many steps are needed before it can be applied in large scale, although it is used in some cases (e.g. the Alpha Magnetic Spectrometer, a superconducting magnet on ISS, detects and deflects cosmic particles \cite{Battiston:2012}). Given that, we do not examine further this strategy in this work. An interested reader may refer to \cite{Marco:2021} for a more thorough analysis of its possible implementation on Mars.

We focus our analysis on the other type of shielding, the passive shielding. In this case, the shielding material fulfils three purposes. Firstly, it has to stop highly energetic particles inside its volume. Secondly, it must minimize the formation of secondary particles resulting from the interaction with the shielding material itself (target fragments). Thirdly, it needs to induce nuclear fragmentation of the incoming projectile. This process is beneficial for protecting astronauts, as it breaks down highly damaging heavy ions into lighter, less harmful fragments. Since the radiation risk comes mainly from the energy lost by incoming particles within a tissue or organ, materials are characterized by their stopping power, with respect to their passive shielding properties. The stopping power is practically the incoming energy lost per unit of thickness through a material. The Bethe-Bloch formula calculates the electronic stopping power $-dE/dx$ of a material (density $\rho$) when a charged particle (charge $z$ in units of electron's charge) travels through its mass \cite{Leo:2012,Durante:2014}: 

%

\begin{multline} \label{1}
   - \frac{dE}{dx} = \frac{4 \pi N_A e^4}{m_e} \rho \frac{Z}{A} \frac{z^2}{\upsilon^2} 
   \left[ \ln \left( \frac{2 m_e \upsilon^2 W_{max}}{I (1 - \beta^2)} \right) \right. \\
   \left. - \beta^2 - \frac{C(\beta)}{z} + Z L_1(\beta) + z^2 L_2(\beta) + L_3(\beta) \right]
\end{multline}

where $N_A$ is Avogadro's number, $e$ and $m_e$ are the electron's charge and mass, $Z$ and $A$ are the atomic number and weight of the material, $\beta = \upsilon /c$ is the normalized velocity of the projectile ($\upsilon$ is the projectile's velocity and $c$ is the speed of light), $W_{max}$ is the maximum energy transfer in a single collision and $I$ stands for the mean excitation potential. The last terms are correction terms. $C(\beta)$ is the shell correction, $L_1(\beta)$ is the Barkas correction, $L_2(\beta)$ is the Bloch term and $L_3(\beta)$ is the Matt and density correction. The gradient of energy deposited through a material is caused by inelastic collisions with the target electrons. More precisely, the quantity often used in literature, linear energy transfer (LET), is defined as the average (radiation) energy deposited to the material electrons per unit path length along the track of an ionizing particle through electronic and nuclear interactions \cite{Durante:2014}.

A crucial parameter in the above expression is $z^2$, which arises from Coulomb's law, and suggests that as the charge of the particle increases, the interaction with the electrons in the material becomes stronger, leading to greater energy loss. The stopping power also scales as 1/$\beta^2$ implying that a shielding material can potentially be counterproductive if it just slows down the incoming radiation, as then the LET would be higher at the end of the shield. In other words, slower particles may be more dangerous than faster ones, which brings up the importance of nuclear fragmentation when all incoming radiation cannot be effectively stopped. 

Likewise, we conclude the stopping power of a material is proportional to $\rho Z/A$. This condition is fulfilled by high-density substances made from chemical elements of low atomic number. It follows from the fact that materials with many electrons per unit of mass, little mean excitation energy and least tight binding corrections are the best energy absorbers. Typical such materials are the ones rich in hydrogen, like water (H$_2$O) and polyethylene (CH$_2$) \cite{Miller:2003,Zeitlin:2008,Guetersloh:2006,Barthe:l2018}. Especially, liquid hydrogen is assumed to offer great protection, but due to poor mechanical properties and chemical instability of materials like that, aluminum is at present most commonly used in aerospace. Note that no material effectively stops cosmic radiation, because of its very high energy and the volume and mass constraints of spacecrafts. The engineering strategy is to take advantage of sudden drops of penetrating energy for a given material and then try to mitigate exposure further using other countermeasures. 

Yet, passive shielding has a critical limitation: additional weights targeted for shielding purposes increase the payload of the launched vehicle \cite{Slaba:2017}. More massive spacecraft need substantial amounts of energy to escape the gravity of Earth and travel through space. So, it's impossible to limitlessly increase the amount of shielding materials (concerning volume constraints) until they provide adequate radiation protection. Therefore, other alternatives are required too, utilizing the in-situ resources and structures. Some suggested ideas are described in the next section. 

\subsection{Other strategies}
In general, shielding strategies are different for the surface of Mars and for deep space transits (or for Moon missions). For instance, while in deep space, protons are the radiation type which dominates behind thick shields \cite{Norbury:2020}, neutrons and gamma rays are also important in the Martian environment due to interactions with the atmosphere and soil \cite{Clowdsley:2001,Le:2009}. So, it is logical to take advantage of the unique features of each specific radiation environment and plan our strategy accordingly. 

An alternative is using terrain features (like craters, cliffs, caves or lava tubes) as protective shields. This option relies on the fact that since rays arrive isotropically, covering a part of sky reduces some percent of these rays, thus reducing accumulated radiation doses. Doses are for example lower inside craters,  contrary to the top of mountains. Similarly, when located up against a cliff, the majority of radiation comes from the diametrically opposite side of the sky \cite{Simonsen:1990,Ehresmann:2021}.

Water-rich regolith acts as a great attenuator against neutrons of energies less than 10 MeV, e.g. \cite{Rostel:2020}. These neutrons undergo elastic scattering on the hydrogen atoms in water or ice, get captured by the latter and are less able to continue their journey or get back-scattered. This is an indirect method of reducing albedo particles doses. Secondary neutrons of energies from 1 to $\sim$100 MeV are of significant biological importance, due to their ability to penetrate deeply into matter \cite{Mountford:1992}. Their biological quality factor (see Appendix) is $Q>20$, as defined by the International Commision on Radiological Protection \cite{Petoussi:2010}, whereas for other particles like protons it is around $Q=1$. For this reason, protecting from neutrons is a critical issue and should never be neglected. Regions rich in subsurface or surface ice (e.g. polar or subpolar regions) are then proposed as landing sites \cite{Rostel:2020}.

Moreover, the employment of local regolith has been suggested in literature (e.g. \cite{Arnhof:2016}), whereas we note that such strategy has been also suggested to be used on the Moon, and in that case, both simulation \cite{Horst:2022} and experimental studies \cite{Luoni:2022} have been made. Unconsolidated regolith may act as the basis to manufacture bricks. These bricks, combined with some kind of epoxy, would constitute a habitat. Other scenarios include piling regolith over dwelling \cite{Ortiz:2015} and tunnelling (digging the subsurface to create protected living places) \cite{Wilson:1997}. Although these approaches are interesting, further issues must be addressed, such as the carry of the mechanical tools required for digging and construction. En route stops on objects like the Moon to gather regolith seem impractical for the time being.

Astronauts themselves should also be protected by adopting personal protection measures. That is, astronauts must limit time spent outside of the habitat, during walks or expeditions. Even when located inside the vehicle, they should stay as protected as possible. Storm shelters, for instance, are places inside the habitat which offer increased protection from radiation. They are useful during solar events, when the magnitude of Sun-originated protons rises enormously. The prompt forecast of the space weather \cite{Baker:2005} would be useful in that case, but it is still an open issue to model the solar activity precisely.

\begin{table*}[h]
    \caption{Adopted composition of Martian atmosphere}    
    \centering
    \begin{tabular}{ccc}
            \hline
            Chemical Compound & Molecular Percentage \\  
            \hline
            CO$_2$ & 95.10 \\ 
            N$_2$ & 02.59 \\
            Ar & 01.94 \\
            O & 00.16 \\
            CO & 00.06 \\
            H$_2$O & 00.15 \\
            \hline 
            Based on data taken by Mars Fact Sheet \cite{Williams:2004}.
            \end{tabular}

    \label{tab:1}
\end{table*}

\begin{table*}[h]
    \caption{Adopted composition of Martian Soil}    
    \centering
    \begin{tabular}{ccc}
            \hline
            Chemical Compound & Molecular Percentage \\  
            \hline
            SiO$_2$ & 51.2\\ 
            Fe$_2$O$_3$ & 9.3 \\
            H$_2$O & 7.4 \\
            Al$_2$CaK$_2$MgNa$_2$O$_7$ & 32.1 \\
            \hline
            Based on regolith definition of the OLTARIS model \cite{Matthia:2016}.
            \end{tabular}

    \label{tab:2}
\end{table*}

Apparently, radiation protection on Mars is not a straightforward practice. Absorbed doses may shift according to a wide range of (probably unpredictable) factors like location, season, atmospheric conditions, solar activity or others. It is unavoidable to account for as many as possible scenarios and solutions before determining each specification of a human mission to Mars.

Lastly, another suggested countermeasure, not however specifically related to Mars, is the consumption of medical and dietary supplements by the astronauts, in order to mitigate the effects of ionizing radiation. These supplements are technically pills, which help human organism to respond better against radiation exposure (e.g. \cite{Kennedy:2011}). It is then quite understood that radiation protection is not a simple one-dimensional issue, but a whole spectrum of different engineering, operational, and dietary designs are required to ensure the safety of astronauts. 

\section{Method}
Radiation risk derives principally from the energy lost by the particles in a tissue. Particles with both high and low energies pass through tissues and deposit parts of their energy in those. Multiple computational programs are adopted by researchers to calculate energy distributions at various points of a tissue or organ, and to determine the beneficial effect of materials used as shields against radiation.

In this work, we use the Geant4 toolkit for our simulations, developed by CERN \cite{Agostinelli:2003}. Geant4 is a Monte-Carlo numerical model that simulates particle propagation, used for a variety of purposes (e.g. medical, engineering) including space science. Geant4 is specifically applied to model interactions of radiation with matter, since it includes a wide range of physics processes (hadronic and EM interactions, decays etc.), and enables complex geometrical configurations. Besides, \cite{Matthia:2016} has shown good agreement between MSL RAD data \cite{Hassler:2014} and the Geant4 code. Geant4 model, in turn, results in compatible results with other codes. Minor discrepancies between different codes may arise due to alternative model setups, geometrical configurations, tally specifications or other reasons. 

As all charged particles interact with matter in different ways, a variety of physics models has been constructed to describe these interactions. These interactions may be electromagnetic or hadronic elastic and inelastic processes. Apparently, an extremely accurate physics model is anti-correlated with the CPU performance. In other words, fast simulations usually provide imprecise results, and conversely. To maintain computational times under logical limits, without though losing much accuracy, we have to select a solution halfway. Specifically, we use the G4Hadron-PhysicsQGSP\_BERT\_HP model for inelastic hadronic interactions and G4EmStandardPhysics\_option3 for electromagnetic ones. Both of these physics lists have been shown to provide accurate enough results for space and medical purposes \cite{Matthia:2016}. As for other interactions (e.g. elastic hadronic, decay) we adopt the pre-defined models of the Hadr07 example\footnote{https://gitlab.cern.ch/geant4/geant4/-/tree/master/examples/extended/hadronic/Hadr07}. Besides, we also ran test simulations using other models like INCLXX finding no significant change in the results. Detailed specifications for the above models may be found in the Geant4 Physics Reference Manual\footnote{https://ftp.cs.up.ac.za/gentoo/distfiles/9a/PhysicsReferenceManual-4.11.2.pdf}.

\subsection{Geometry}

\begin{table*}[h]
    \caption{Martian geometrical configuration}    
    \centering
    \begin{tabular}{cccc}
            \hline
            Layer & Content & Bulk density (g $\cdot$ cm$^{-3}$) & Thickness (m) \\  
            \hline
            1 & Condensed atmosphere & 0.023 & 10\\ 
            2 & Surface atmosphere & 0.00002 & 1\\
            3 & Shielding material & (varying) & (varying)\\
            4 & Surface atmosphere & 0.00002 & 2 \\
            5 & Water & 1.0 & 1 \\
            6 & Soil & 1.7 & 3\\
            \hline
            \end{tabular}

    \label{tab:3}
\end{table*}

\begin{figure}
    \centering
    \includegraphics[width=0.5\textwidth]{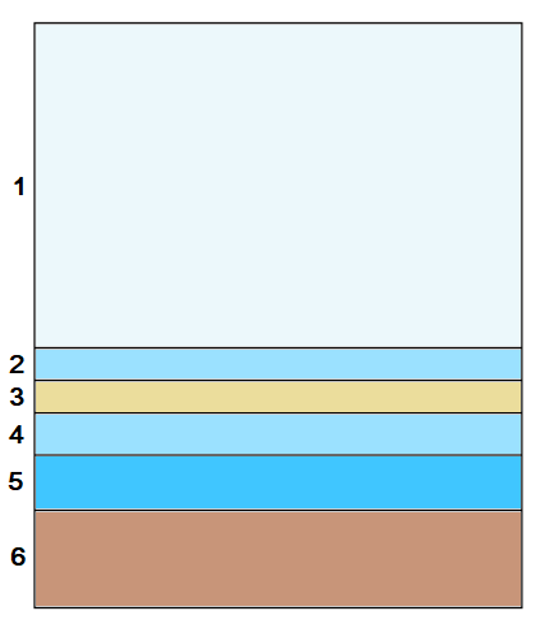}
    \caption{Martian geometrical configuration used for simulations. It consists of 6 successive layers of total length equal to 18 m: Layer 1 (Condensed atmosphere), 2 (Surface atmosphere), 3 (Shielding material), 4 (Surface atmosphere), 5 (Water) and 6 (Soil). Primary radiation hits at the top of the layers.}
    \label{fig:1}
\end{figure}
\begin{figure*}
    \centering
    \includegraphics[width=0.8\textwidth]{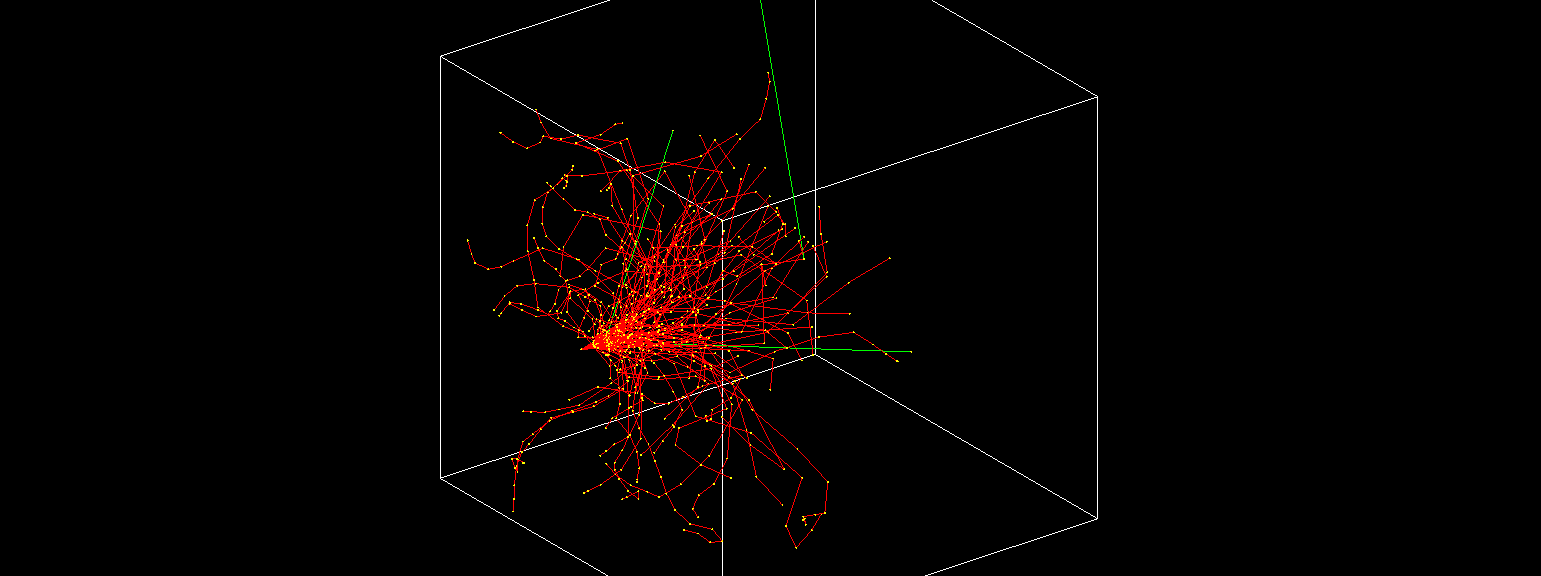}
    \caption{Visualization of the rays' trajectories originating from a pencil beam with a small beam angle.}
    \label{fig:25}
\end{figure*}

Our geometry setup is similar to the one used in \cite{de-Wet:2017}. Fig. \ref{fig:1} depicts the model we constructed to simulate Mars. Adopted Martian atmosphere and regolith compositions used in our simulations are displayed in Tables \ref{tab:1} and \ref{tab:2} respectively. We used the Martian Fact Sheet data in the first case \cite{Williams:2004} and the definition of the OLTARIS model in the latter \cite{Matthia:2016}. Our model is set to compute radiation about 4 km below the mean Martian surface, at which depth the RAD instrument is located. In that way, we were able to compare and validate our results.

Our simulation world, as it is called in the Geant4 documentation, is a cubic box which consists of 6 successive layers. Each layer is considered homogeneous in composition and density. At first, a 10-m-thick layer of the condensed atmosphere is defined. We rescaled the actual size (about 100 km) of the atmospheric layer by compressing it and assigning its density accordingly, so that its linear thickness remains constant. The second layer of the simulation world is also an atmospheric layer, now having its normal density and a maximum length of 2 m. Although realistically an astronaut would need Earth air pressure in his habitat, this thin layer does not really matter at Martian pressure. The tested shielding material comes next, which is followed by another surface atmosphere layer standing for the atmosphere within the habitat. The fifth layer is a water slab, which represents an astronaut located inside a habitat on Mars. A final layer of regolith (thickness of 3 m), which acts as a collector for particles not absorbed or reflected in previous layers, closes the simulation box. Table \ref{tab:3} summarizes the specifications of each layer. 

In such a way, Mars is represented as series of absorbing layers, spanning from the top of its atmosphere (about 100 km above mean zero elevation) up until some meters below its surface. The layer 3, as above described, represents the roof of a habitat. We test several potential materials of which it may be constructed, and change the respective thicknesses as well, in order to figure out how radiation responds to varying lengths of shield. As for a human phantom, we use the water slab, since an average tissue is made almost completely out of water. The two left dimensions of the simulation world (x and y) are large enough to avoid the escape of particles, which could falsely reduce the calculated doses. All particles, including recoil protons, are confined to the simulation world (other ions are not affected so much anyways because of their higher stopping rate, $z^2$).

We acknowledge that the adopted geometry is quite simplified. For example, we could have defined multiple atmospheric layers to account for density and temperature variations through altitude. However, setting an extremely detailed model of Mars is beyond the scope of this study, since our aim is to qualitatively distinguish materials in accordance to their shielding properties. For this purpose, a model like the one we adopted will suffice.

\begin{figure*}
    \centering
    \includegraphics[width=0.8\textwidth]{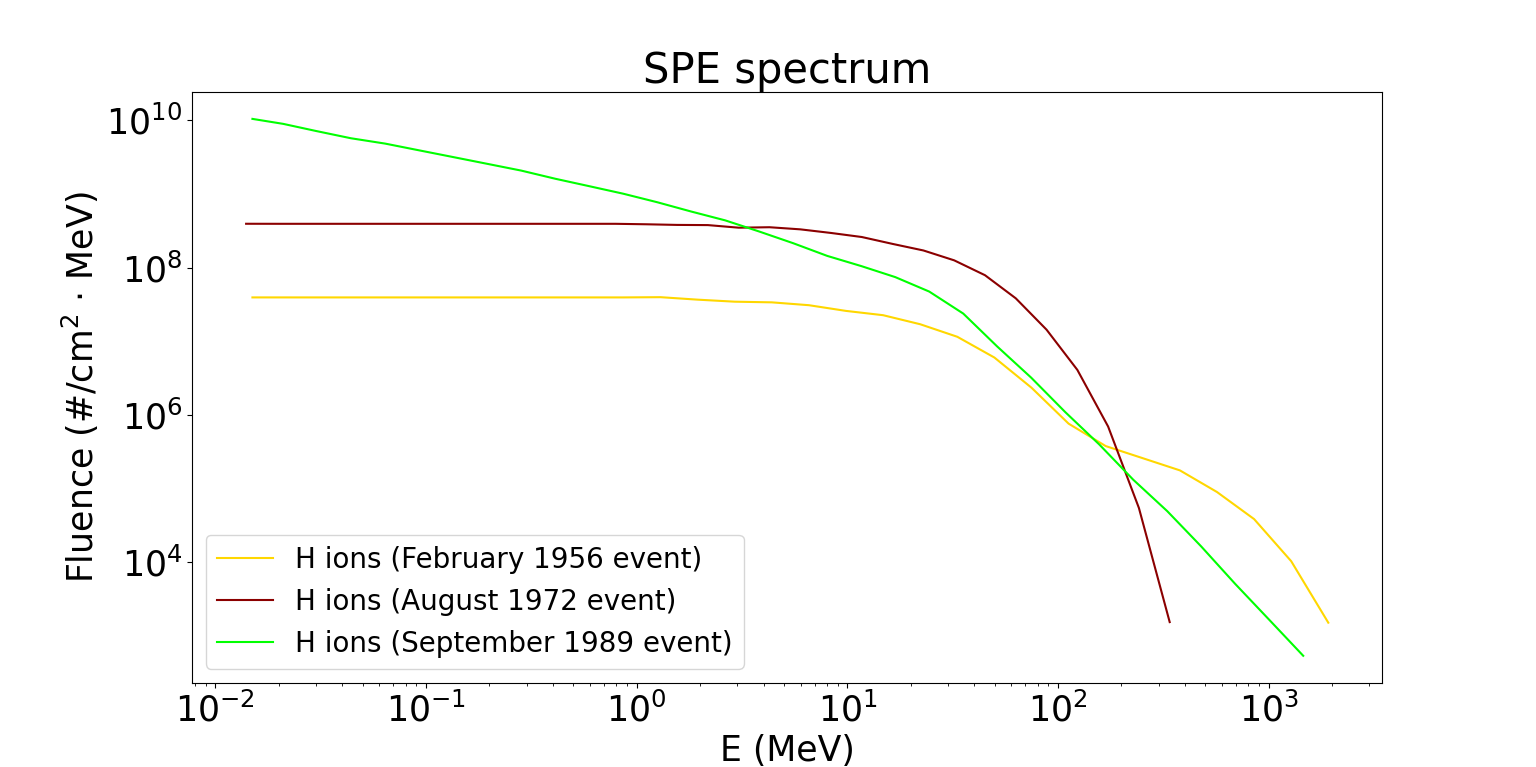}
    \caption{The SPE proton spectra during the events of February 1956, August 1972 and September 1989. Based on data taken by \cite{Clowdsley:2006} and corrected by the 1/r$^2$ dependence of  SPE fluence}.
    \label{fig:2}
\end{figure*}
\begin{figure*}
    \centering
    \includegraphics[width=0.8\textwidth]{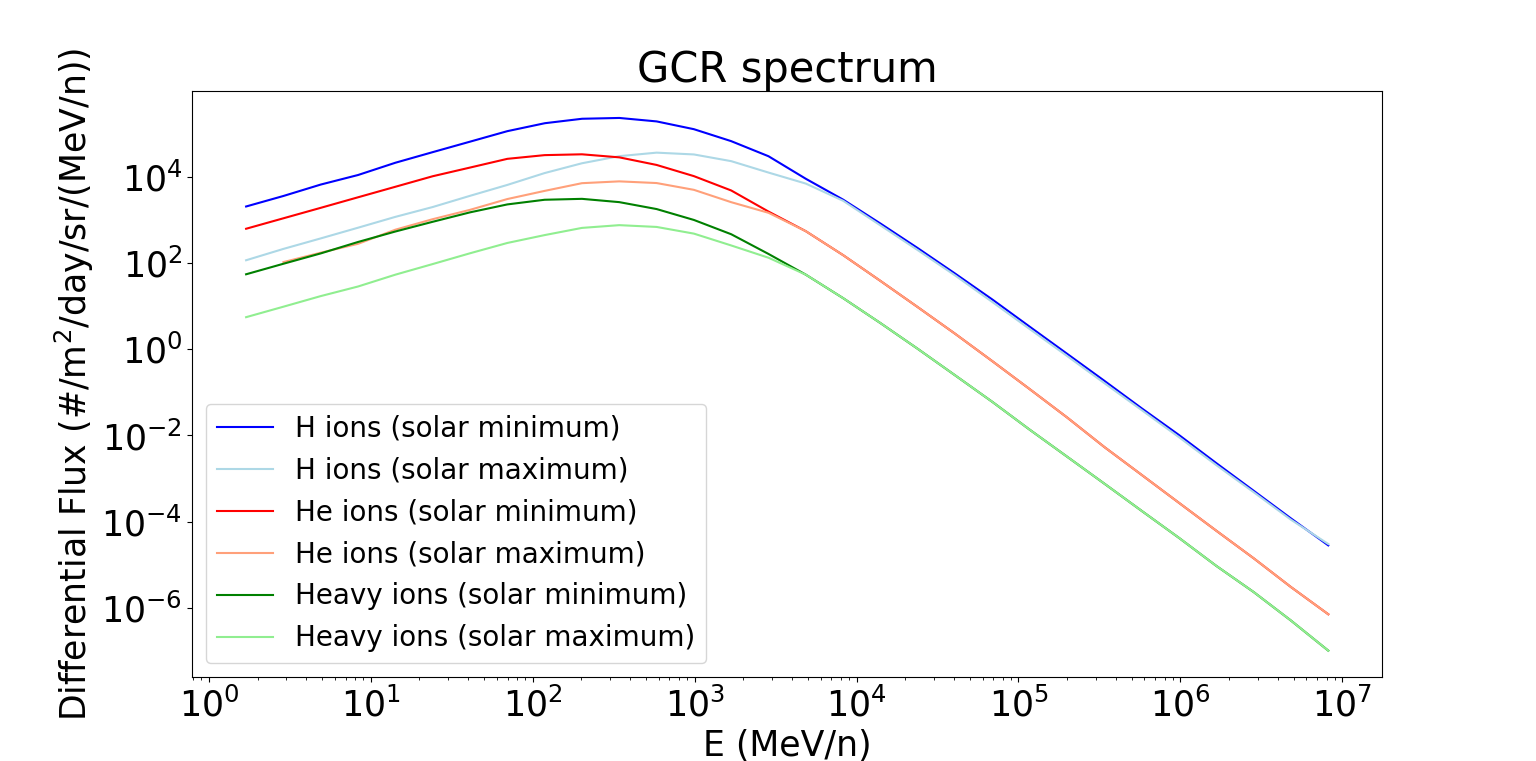}
    \caption{The differential spectrum of GCR, as calculated by \cite{Neill:2010}. Blue lines stand for protons, red for alpha particles and green for heavier nuclei. Bright colors represent the conditions of minimum solar modulation and lighter ones for maximum solar modulation conditions.}
    \label{fig:3}
\end{figure*}

\subsection{Primary particles}
As mentioned previously, the Geant4 model has been demonstrated to simulate particle spectra on the surface of Mars accurately enough for many particles, such as protons, alpha particles or heavy ions, as verified by the MSL RAD instrument \cite{Matthia:2016}. However, results concerning other types of particles, like neutrons, are not that close. 

We choose to compare the stopping power for a stream of many particles rather than the behavior of a material on one particle solely. In that way, it is ensured that there are not any random interactions between the material and the particle, affecting the net shielding effect, but all arbitrary behavior is averaged out. More precisely, an almost pencil beam (the angular distribution is 0.003$^{\circ}$) consisting of 1,000,000 primary particles for SPEs and 100,000 ones for GCR hits the model at the top. 
The statistical error of the simulations scales to $1/\sqrt{N}$, where N is the number of particles (primaries and those reaching the in-question depths), therefore, the error in the results is roughly 0.1-0.3\%, which allows us for robust conclusions. A point source generates a flux of the primary particles, downward directed, and as shown in Fig. \ref{fig:25}, rays gradually cover the entire simulation world and travel through every direction traversing various paths through the layers, while albedo particles are included as well (hence, doses are summed up and calculated accounting for the entire water layer). The reason for this choice was to minimise the escape of particles outside of the simulation world; that is why a non-isotropic initial beam was tested, apart from reasons related to computational times. 

We analyze the response of shielding materials against cosmic radiation coming from both sources, as defined in Section 2.1 (SPEs and GCR). In the first case, we use three major historic solar events as representatives: February 1956, August 1972 and September 1989. As solar eruptions show notable differences with each other in terms of intensity, abundance, composition and duration, different input events of them are required to better study their effect. Fig. \ref{fig:2} depicts the energy spectrum of solar protons produced by these three events. Data were obtained by \cite{Clowdsley:2006} and were corrected according to the 1/r$^2$ dependence of SPE fuence (they were divided by 1.5$^2$). On the other hand, the Badhwar-O’Neil 2010 model (\cite{Neill:2010}) is used to obtain the background GCR spectrum. Protons, alpha particles and iron nuclei (as a substitute for heavier particles) are considered. These particles are chosen based on their dominance in the GCR composition and their significant impact on the radiation dose. Including the full spectrum of GCR particles would significantly increase the complexity and computational load of the simulations. Using iron as a representative for heavy species saves computational time and resources, making the analysis feasible without overwhelming computational capacity. The primary GCR spectrum used in our simulations is shown in Fig. \ref{fig:3}. 

\section{Results}

\begin{figure*}
    \centering
    \includegraphics[width=0.8\textwidth]{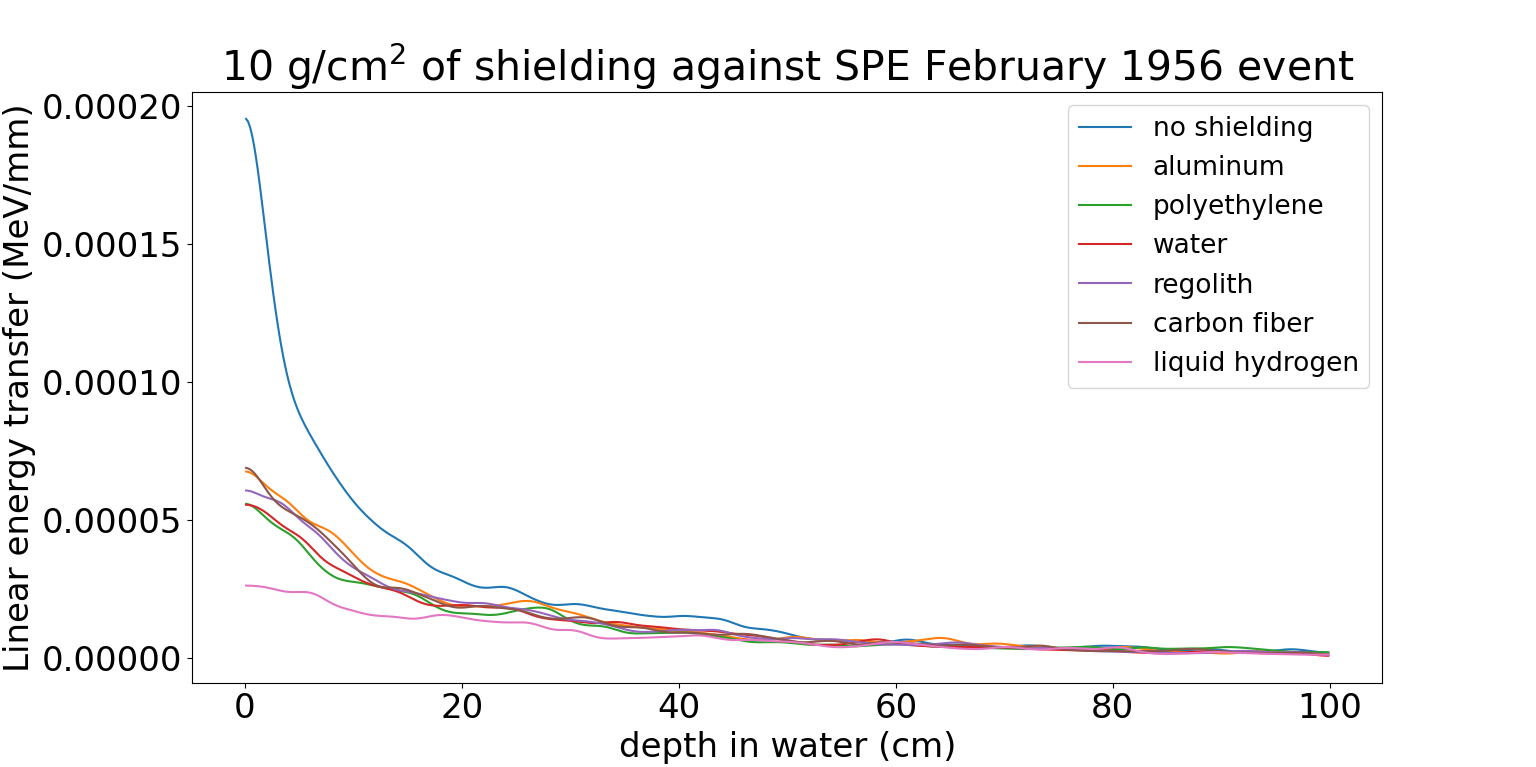}
    \caption{LET for the February 1956 solar event versus depth in the water layer for no shielding material, aluminum, polyethylene, water, regolith, carbon fiber and liquid hydrogen. All of the materials have an area density of 10 g cm$^{-2}$.}
    \label{fig:4}
\end{figure*}

\begin{figure*}
    \centering
    \includegraphics[width=0.8\textwidth]{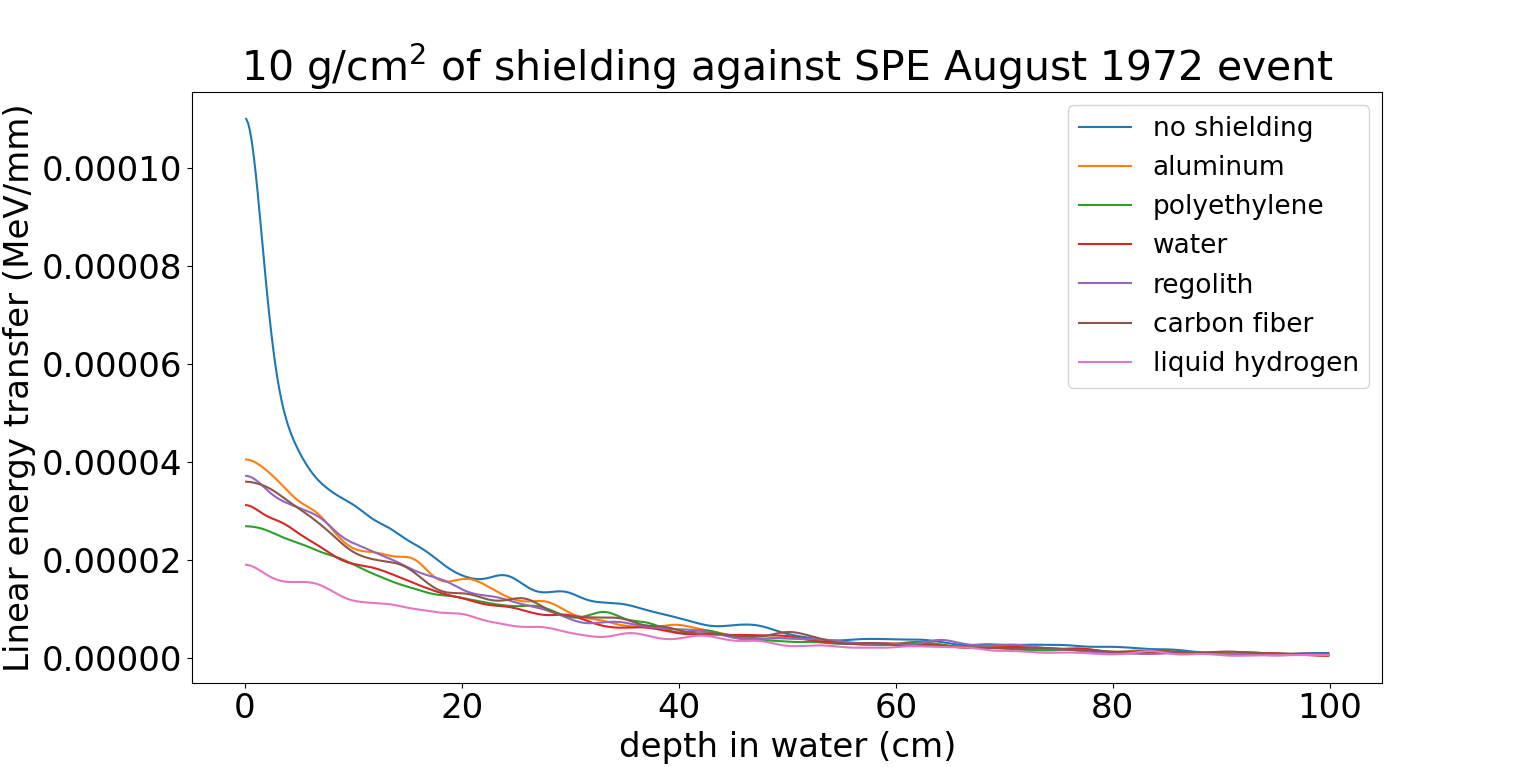}
    \caption{LET for the August 1972 solar event versus depth in the water layer for no shielding material, aluminum, polyethylene, water, regolith, carbon fiber and liquid hydrogen. All of the materials have an area density of 10 g cm$^{-2}$.}
    \label{fig:10}
\end{figure*}

\begin{figure*}
    \centering
    \includegraphics[width=0.8\textwidth]{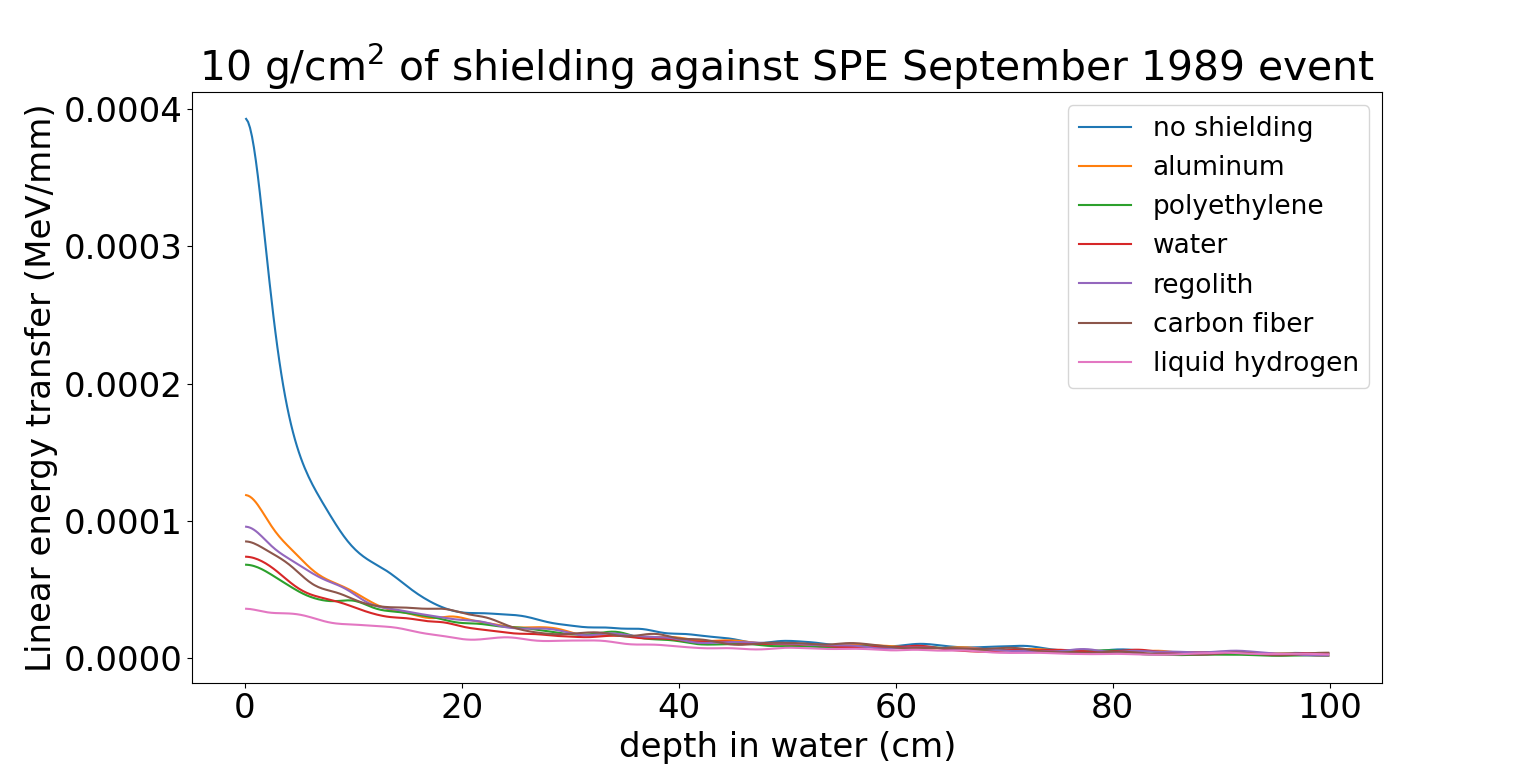}
    \caption{LET for the September 1989 solar event versus depth in the water layer for no shielding material, aluminum, polyethylene, water, regolith, carbon fiber and liquid hydrogen. All of the materials have an area density of 10 g cm$^{-2}$.}
    \label{fig:11}
\end{figure*}

\begin{figure*}
    \centering
    \includegraphics[width=0.8\textwidth]{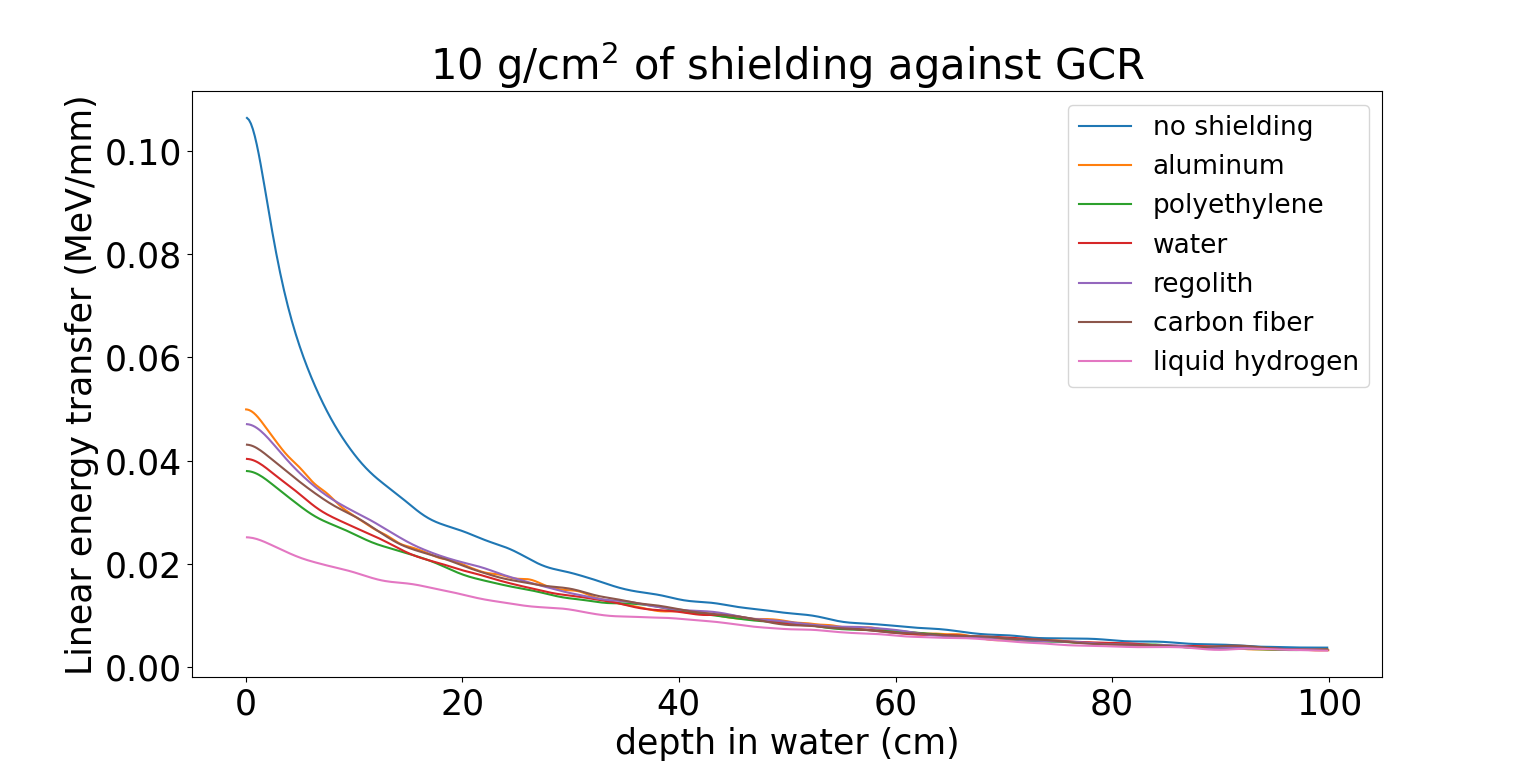}
    \caption{LET by the GCR particles (minimum solar conditions) versus depth in the water layer for no shielding material, aluminum, polyethylene, water, regolith, carbon fiber and liquid hydrogen. All of the materials have an area density of 10 g cm$^{-2}$.}
    \label{fig:5}
\end{figure*}

\begin{figure*}
    \centering
    \includegraphics[width=0.8\textwidth]{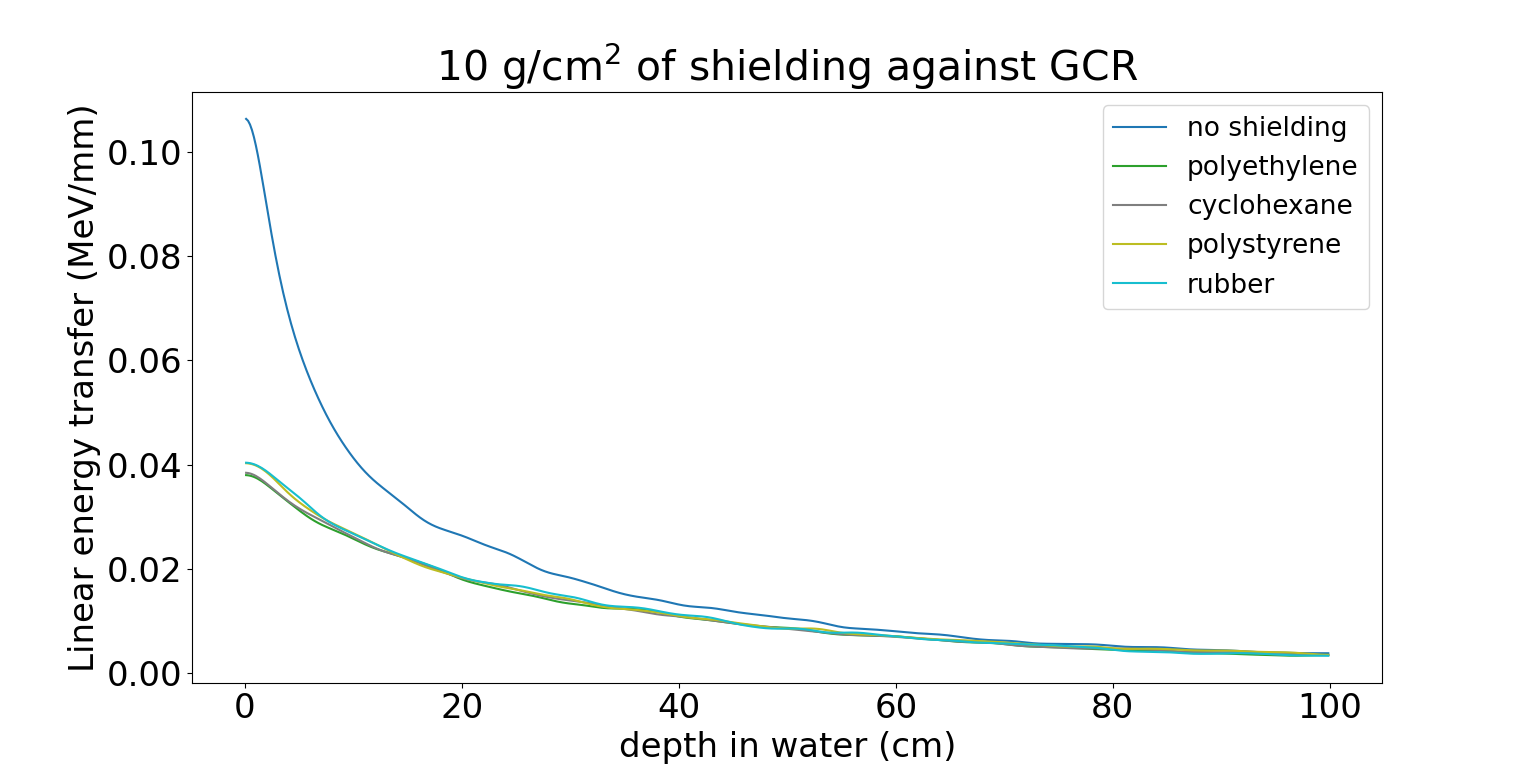}
    \caption{LET by the GCR particles (minimum solar conditions) versus depth in the water layer for no shielding material and the hydrocarbon materials (polyethylene, cyclohexane, polystyrene and rubber). All of the materials have an area density of 10 g cm$^{-2}$.}
    \label{fig:6}
\end{figure*}

As a first step, we ran the fundamental situation of our simulations, i.e. the situation without any shielding material. Hence, radiation could travel freely through the atmospheric layers before reaching the layer representing the astronaut's body, mimicking the actual conditions and natural shielding of Mars. The resulting dose rate, induced by GCR, ranged in our model between 172.8 $\mu$Gy/day (maximum solar conditions) and 296.7 $\mu$Gy/day (minimum solar conditions). These values are quite consistent with the RAD measurements in the plastic detector, which lie in the interval of $\sim$ 200 - 325 $\mu$Gy/day from MSL landing until late 2020 \cite{Guo:2021}. This consistency is particularly encouraging considering the simplicity of the model we built, and shows that our outcomes should be compatible with other outcomes deduced by much more complex models.

We continue with examining the effect of the shielding materials. The quantity we compute to first-order compare them is energy deposited per depth (LET) versus the depth in the water layer (a more definite classification is given afterwards, using the resulting dose rates for each material). Although the definition of LET usually refers to a monoenergetic single-ion source, in this paper we generalize its computation, so that it is calculated at a given depth by integrating the differential energy loss of all particles over the energy spectrum. While individual particles exhibit a Bragg peak behavior, the spectrum-averaged LET can decrease with depth due to the cumulative effects of all particles, including those that have already deposited their energy and stopped.

The LET versus the depth in the water layer, caused by three typical solar events, are shown in Fig. \ref{fig:4}, \ref{fig:10} and \ref{fig:11} for some specific materials, commonly used or proposed in aerospace. The materials tested are aluminum, polyethylene, water, regolith, carbon fiber and liquid hydrogen, and they are assigned the same area thickness, 10 g cm$^{-2}$. That way, we examine how helpful against cosmic rays a material is, accounting simultaneously for its weight efficiency, as all of the simulated cases would have the same mass. For comparison, the situation with no shielding material is given too. In Fig. \ref{fig:5}, the behavior of the same materials against the full GCR spectrum (protons, alpha particles and helium ions) during solar minimum conditions (as a worst case scenario) is presented. 

\begin{table}
    \caption{Dose rates (from highest to lowest) in the water layer induced by GCR particles (minimum solar conditions) for shielding materials (10 g cm$^{-2}$ of thickness).}  
    \centering 
    \begin{tabular}{lc}
            \hline
            Material & Dose rate ($\mu$Gy/day) \\  
            \hline
            no shielding & 296.7 \\ 
            aluminum & 214.3 \\
            regolith & 212.6  \\
            BN & 212.6 \\
            beryllium & 212.6 \\
            SiO$_2$ & 209.3 \\
            carbon fiber & 207.7 \\
            ETFE & 207.7 \\
            Mylar & 204.4 \\
            rubber & 201.1  \\
            Kevlar & 201.1 \\
            water & 199.4  \\
            PMMA & 199.4 \\
            polystyrene & 199.4 \\
            LiH & 196.1 \\
            cyclohexane & 194.5  \\
            polethylene & 192.9 \\
            LiBH$_4$ & 189.6 \\
            liquid hydrogen & 152.0  \\
            \hline
            \end{tabular}

    \label{tab:4}
\end{table}

\begin{table*}[h]
    \caption{Comparison of the dose rates ($\mu$Gy/day) in the water layer induced by GCR particles (minimum solar conditions) for shielding materials and their combinations with regolith}    
    \centering 
    \begin{tabular}{lcc}
            \hline
            Material & 10 g cm$^{-2}$ material & 5 g cm$^{-2}$ material, 5 g cm$^{-2}$ regolith\\  
            \hline
            aluminum & 214.3 & 215.9\\ 
            carbon fiber & 207.7  & 212.6\\
            water & 199.4 & 204.4\\
            polyethylene & 192.9 & 201.1\\
            \hline
            \end{tabular}

    \label{tab:5}
\end{table*}

\begin{table*}[h]
    \caption{Comparison of the dose rates ($\mu$Gy/day) in the water layer induced by GCR particles (minimum solar conditions) for shielding materials and their combinations with aluminum}    
    \centering 
    \begin{tabular}{lcc}
            \hline
            Material & 10 g cm$^{-2}$ material & 5 g cm$^{-2}$ material, 5 g cm$^{-2}$ aluminum\\  
            \hline
            carbon fiber & 207.7 & 214.3 \\
            regolith & 212.6 & 212.6\\
            water & 199.4 & 206.0\\
            polyethylene & 192.9 & 202.7\\
            \hline
            \end{tabular}

    \label{tab:6}
\end{table*}

Energies deposited are definitely lower in the SPE-induced diagrams than the GCR-induced ones, since the latter are characterized by a much harder spectrum (though we note that slower particles can have higher damaging effect, as pointed earlier). 
In spite of the low energies, some basic results may be inferred. For instance, aluminum seems in all cases to be the worst shielding material, among the ones tested, or, liquid hydrogen indeed maintains the exposure levels at the lowest levels, as expected by eq. (1). Only minor differences are observed in the response of the materials against each one of the three different solar events. Next, we focus on the GCR particles, which form the main field on Mars, as explained in Section 2.2. 

We note that results by the GCR ions (Fig. \ref{fig:5}) are compatible with the SPEs-related ones in Fig. \ref{fig:4}-\ref{fig:11}. Again, liquid hydrogen poses as the best attenuator, since it reduces roughly almost to half the absorbed doses in comparison with the other materials. The rest of them have an intermediate behavior, reducing radiation to about a third of the situation of no shielding at all. Our results are in agreement with \cite{Ortiz:2015,Durante:2014}, which also simulate the passive shielding performance on the Martian field. In order of energy decrease, materials are ranked as: liquid hydrogen, polyethylene, water, carbon fiber, regolith, aluminum. Even the differences between the curves of the simulated materials in Fig. \ref{fig:5} are almost identical with the respective differences in the plots of Fig. 3 of \cite{Ortiz:2015}. As mentioned above, the statistical error in the simulations is about 0.3\%, and thus, we can confidently assume that these deviations (at minimum $\sim$1-2\%) do not come from the random nature of the Monte Carlo calculations, but illustrate the particular behavior of each material against radiation instead.

We now turn our attention on regolith's performance. As mentioned, regolith could significantly assist on the radiation shielding purposes, as it would reduce the required payload. Thus, assessing the extent at which Martian regolith could fulfil this purpose is an important issue. Our results suggest that its behavior against charged particles is similar to aluminum's (it is slightly more effective). We remind that the same thickness in g cm$^{-2}$ is simulated here. This outcome is also supported by experimental results of the Moon's regolith \cite{Luoni:2022}, which has a similar composition with the Martian one. Other materials, like the hydrogen-based ones, diminish absorbed doses more, but the difference is only small (though definitely larger than the statistical uncertainties). Consequently, regolith is classified as a relatively quite effective materials against SPEs and GCR. Previous studies deduce similar outcomes. In particular, \cite{Morthekai:2007,Ortiz:2015,Chen:2022} conclude that $\sim$1 m of regolith shielding provides adequate protection, but increasing the thickness attenuates radiation more gradually; in order to keep effective dose below the safety limits more than 3 m of regolith are required. Mixing regolith with graphite epoxy is proposed by \cite{Kim:1998}, because of the hydrogen compounds epoxy contains.

A quantitative comparison of the shielding effects of each one of materials can be found in Table \ref{tab:4}, where we present the GCR-induced (minimum solar conditions) dose rates (see Appendix) in the water layer (doses are calculated from the total energy deposited in this slab). As noted above, the resulting doses for the hydrogen-based ones are very close to each other, while liquid hydrogen shielding results in by far the lowest dose (51\% of the dose without passive shielding).

We further examine the performance of some materials, similar to polyethylene, as well. Precisely, these materials composed out of carbon and hydrogen only, are: high-density polyethylene ((C$_2$H$_4$)$_n$), cyclohexane (C$_6$H$_{12}$), polystyrene ((C$_8$H$_8$)$_n$) and styrene butadiene (C$_{20}$H$_{22}$)$_n$), as a typical family of rubber. All of these materials could be useful in aerospace missions due to their properties and lightness. For example, polyethylene and polystyrene are two of the most widely used plastics. Their plots are given in Fig. \ref{fig:6}. We conclude that they respond to radiation almost identically, and small differences are noticed. This is of course expected, due to their analogous chemical identity. As it can be inferred from Table \ref{tab:4}, there is a slight increase in the dose rate as the number of carbon atoms rises, hence in order of increasing dose deposition the materials are ranked as: polyethylene, cyclohexane, polystyrene and lastly, rubber.

\begin{figure*}
    \centering
    \includegraphics[width=0.8\textwidth]{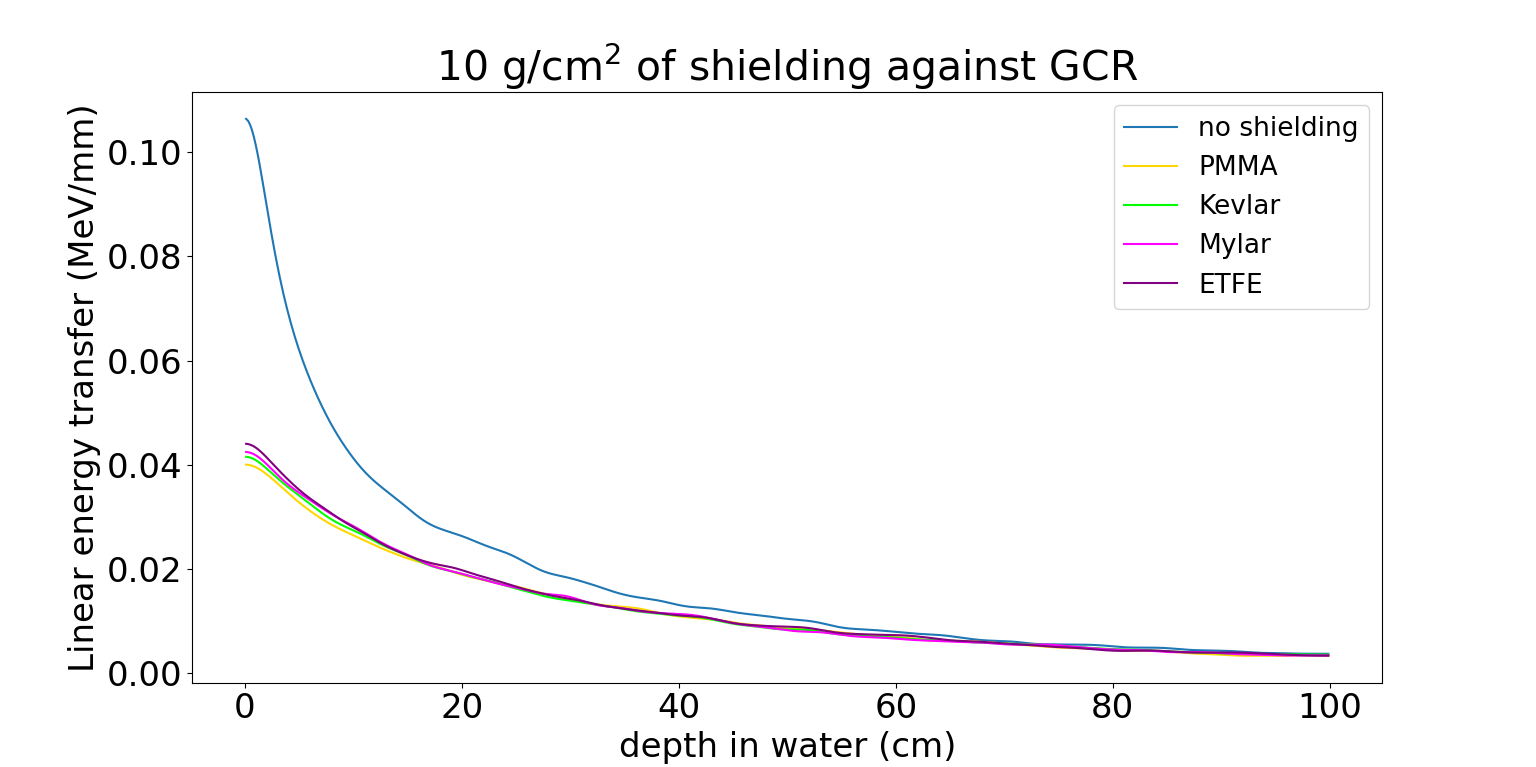}
    \caption{LET by the GCR particles (minimum solar conditions) versus depth in the water layer for no shielding material, PMMA, Kevlar, Mylar and ETFE. All of the materials have an area density of 10 g cm$^{-2}$.}
    \label{fig:15}
\end{figure*}

\begin{figure*}
    \centering
    \includegraphics[width=0.8\textwidth]{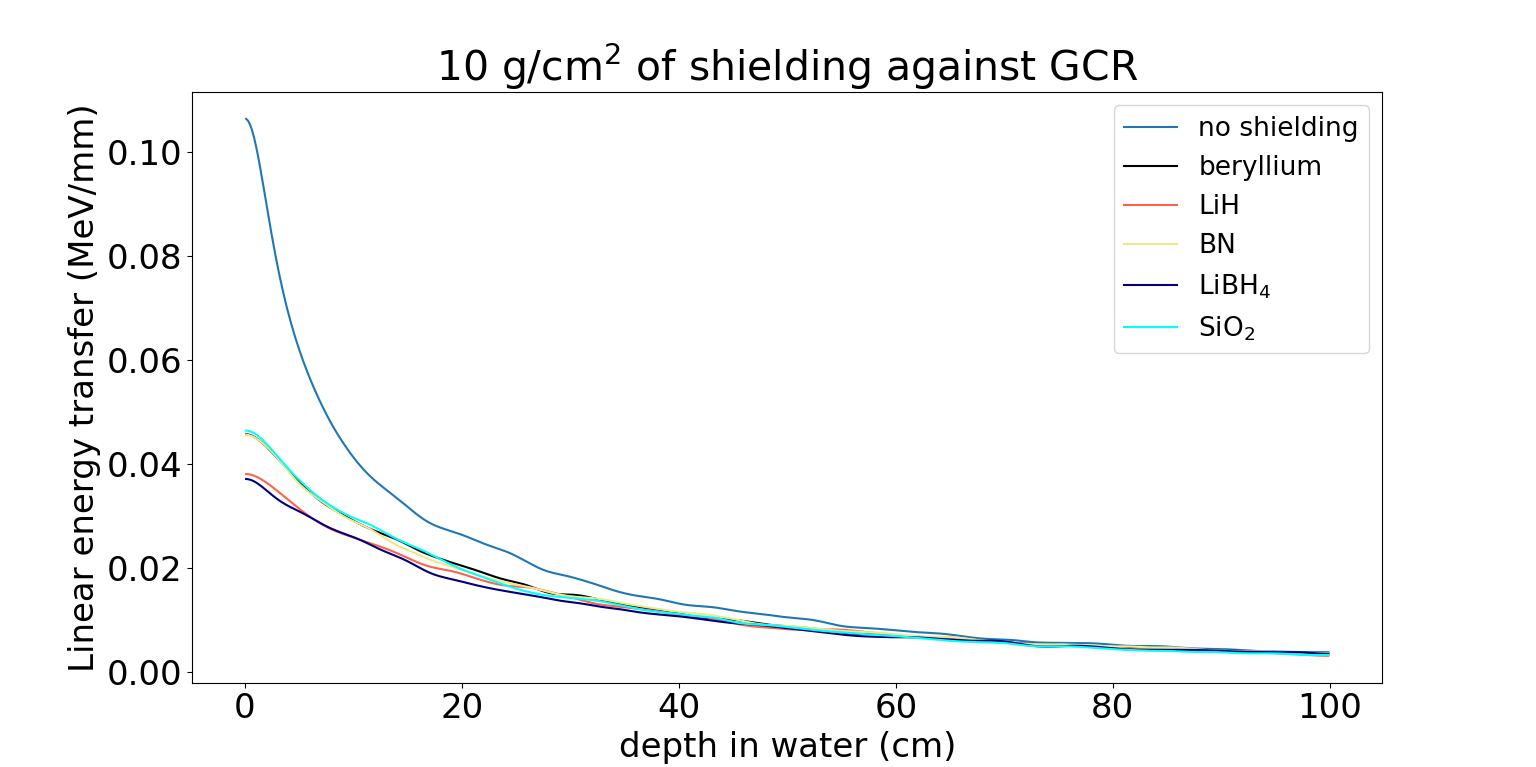}
    \caption{LET by the GCR particles (minimum solar conditions) versus depth in the water layer for no shielding material, beryllium, LiH, BN, LiBH$_4$ and SiO$_2$. All of the materials have an area density of 10 g cm$^{-2}$.}
    \label{fig:16}
\end{figure*}

Fig. \ref{fig:15} depicts a comparison of the shielding properties among organic compounds again based hydrogen, but combined with other elements like nitrogen, oxygen and fluorine. We test PMMA ((C$_5$O$_2$H$_8$)$_n$), Kevlar (C$_{14}$H$_{14}$N$_2$O$_4$)$_n$), Mylar (C$_{10}$H$_8$O$_4$)$_n$) and ETFE (C$_{4}$H$_{4}$F$_4$)$_n$). The respective plots and dose rates show again a similar performance, but lower-density PMMA seems to be more effective than the densest ETFE. Kevlar and Mylar result in similar doses.

As explained in Section 3.1, low-atomic mass materials tend to be better attenuators against any type of radiation. Motivated by this opinion, which is strongly supported by previous analyses, we simulate a series of innovative low-Z chemical compounds. For example, deep-space simulations using GCR spectra (e.g. \cite{Pawlicki:2022}) or monoenergetic beams (e.g. \cite{Naito:2020,Sangwan:2022}), as well as experiments (e.g. \cite{Giraudo:2018,Schuy:2019,Luoni:2022}), conclude that materials based on Li, B and N could have promising performances against cosmic radiation. Our results concerning LiH, BN, LiBH$_4$ are presented in Fig. \ref{fig:16}. In the same figure, the plots by Be, a low-atomic mass metal with proven mechanical properties (though problematic due to its toxicity) and by SiO$_2$, which is the basis of glass are given as well. Results indicate that LiH and LiBH$_4$ are quite effective in the Martian environment, and their dose rate is comparable (or even lower) with the other hydrogen-based materials. Particularly, LiBH$_4$ appears slightly more effective than LiH, an outcome apparently in contradiction with the experimental results of \cite{Luoni:2022}, but this discrepancy is properly resolved considering the mixed spectrum we adopt with respect to the monoenergetic iron beam of the experiments. On the other hand, at $\sim$ 210 $\mu$Gy/day (Table \ref{tab:4})) BN, Be and SiO$_2$ are similar to regolith or aluminum in terms of radiation (indeed, SiO$_2$ is the main component of regolith).

\begin{figure*}
    \centering
    \includegraphics[width=0.8\textwidth]{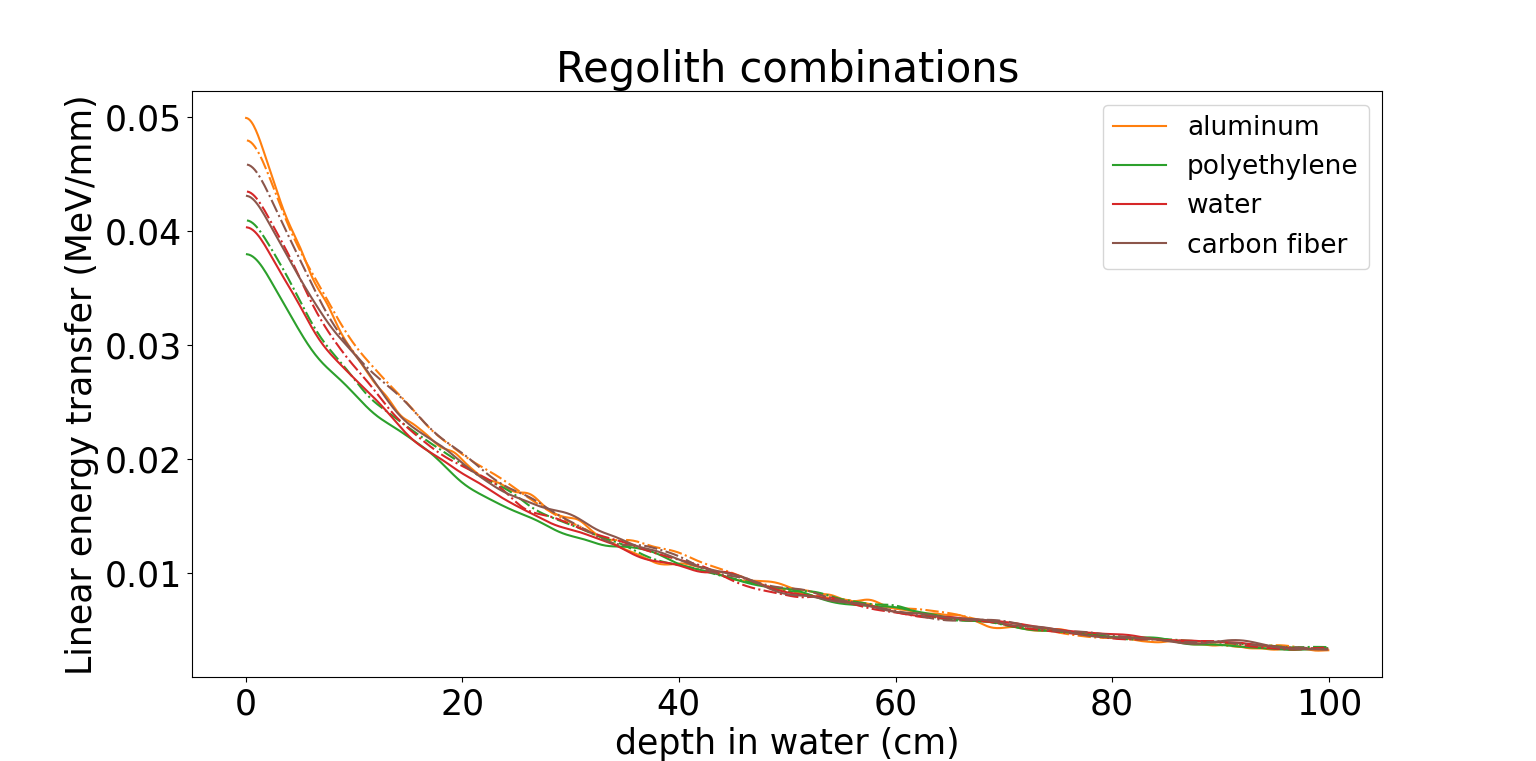}
    \caption{LET by the GCR particles (minimum solar conditions) versus depth in the water layer for combinations of regolith (aluminum, polyethylene, water and carbon fiber). Solid lines represent 10 g cm$^{-2}$ of a material and dash-dotted lines 5 g cm$^{-2}$ of regolith followed by 5 g cm$^{-2}$ of the material.}
    \label{fig:8}
\end{figure*}

Lastly, we simulate combinations of the main passive shielding materials. As a first scenario, we consider how efficient the piling up of regolith on top of the habitat would be, compared to having the material of the habitat solely. Outcomes are demonstrated in Fig. \ref{fig:8} and Table \ref{tab:5}. Solid lines in Fig. \ref{fig:8} represent 10 g cm$^{-2}$ of a material and dash-dotted lines 5 g cm$^{-2}$ of regolith followed by 5 g cm$^{-2}$ of the material. We see that combinations with regolith are less helpful against the charged particles. In spite of this, they still reduce incoming radiation at some considerable extent and differences between the solid and dash-dotted lines in Fig. \ref{fig:8} are only small ($<$ 0.1 MeV/mm) and mainly evident at shallow depths (as expected due to different stopping powers). Regolith properties are actually comparable with the aluminum's, which means that the obtained plot of the sequence 5 g cm$^{-2}$ of regolith and aluminum is quite similar with the one coming from 10 g cm$^{-2}$ of aluminum. This is another confirmation of the similar performance between regolith and aluminum. These qualitative considerations are also supported by the dose rates in Table \ref{tab:5}, where it is shown that the combinations of regolith allow slightly larger dose rates.

\begin{figure*}
    \centering
    \includegraphics[width=0.8\textwidth]{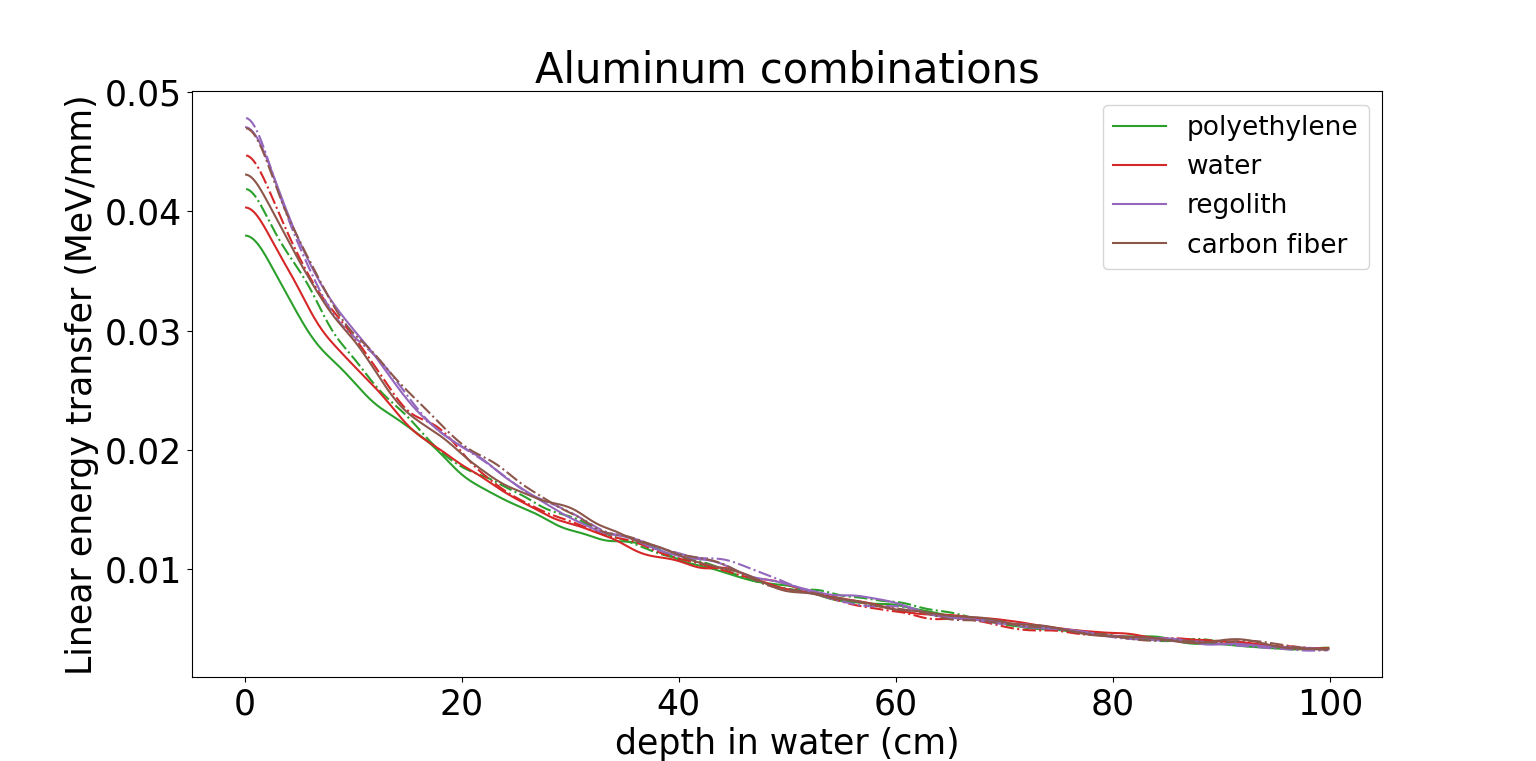}
    \caption{LET by the GCR particles (minimum solar conditions) versus depth in the water layer for combinations of aluminum (polyethylene, water, regolith and carbon fiber). Solid lines represent 10 g cm$^{-2}$ of a material and dash-dotted lines 5 g cm$^{-2}$ of the material followed by 5 g cm$^{-2}$ of aluminum.}
    \label{fig:9}
\end{figure*}

Next, we perform similar simulations with aluminum combinations (Fig. \ref{fig:9} and Table \ref{tab:6}). The reason behind these options lies in the fact that the most parts of spacecrafts are currently being constructed with aluminum; its performance against cosmic radiation is then of notable interest. The plots of Fig. \ref{fig:9} indicate that polyethylene, water and carbon fiber mitigate radiation doses more effectively solely, than when replaced with aluminum of the same weight. 5 g cm$^{-2}$ of the regolith followed by 5 g cm$^{-2}$ of aluminum indicate an almost identical behavior with the energy deposition by 10 g cm$^{-2}$ of a regolith, as stated above. We have to mention though, that although aluminum's characteristics result in a bit higher doses, dissimilarities with different materials are not large enough to exclude aluminum from a suitable shielding material
. On the contrary, we propose taking advantage of combinations of aluminum with other material in order to reduce radiation doses, as suggested in Fig. \ref{fig:9} and the values in Table \ref{tab:6}.

Composite materials are the future of radiation passive shielding in space. Apart from hydrogen, other chemical elements could act as basis for combing materials effective against radiation. Lithium is a typical example. They could be added within the reliable aluminum and regolith combinations, and form materials fulfilling any required use and maintaining doses at low levels.

Summarizing, results indicate that dose reduction is a function of many parameters. Assuming the same incident radiation (e.g., a GCR spectrum, as examined here), the composition and properties of each material play crucial roles. While density is important (for instance, high-density materials tend to increase the likelihood of nuclear interactions enhancing the fragmentation process), the primary factors influencing shielding effectiveness are the atomic weight (affecting the nuclear fragmentation cross section) and the ratio of atomic number to atomic weight as derived from the Bethe equation. Materials with large atomic weights based on low-Z atoms tend to be better shields, as expected by eq.~(\ref{1}). The different composition of the materials (term $\rho Z/A$) explains why similar materials show small differences in terms of shielding properties. The energy lost by a particle as it traverses matter is chiefly due to electron density. Ultimately, payload constraints during missions on Mars necessitate the consideration of lighter materials.

\section{Discussion}
Despite its simplicity, our model has been demonstrated to provide a sufficient simulation of the radiation conditions on Mars. We show that our results are compatible with already published ones, either with alternative and more complex models or with the actual measurements by RAD instrument onboard Curiosity rover. Therefore, we infer that our above described results, coming from the computational outputs, have a high degree of confidence. 

Additionally, we note that our main conclusions regarding the efficiency of various materials are typically consistent with what we expect from eq.~(\ref{1}), as explained. Despite that, our Geant4 runs are indeed required to draw firm conclusions. They take into account the complex interactions between particles and materials, including nuclear reactions such as spallation and fragmentation, and electromagnetic processes like ionization and bremsstrahlung, compared to Bethe-Bloch equation which provides an average response and  does not reveal nuances in the interactions (e.g. scattering events, secondary interactions, particle showers and particle decays). Apart from that, our model explores a wide energy range and several complex compositions and structures that cannot be captured by a generalized formula.

Nevertheless, one limitation of our model is related to the emphasis mainly given in ionizing radiation. While the Geant4 model employed in this study has demonstrated accuracy in simulating particle spectra for various ions, its performance regarding non-ionizing radiation and especially neutrons remains a challenge, as has been indicated by discrepancies in comparison to the MSL RAD instrument results (albedo particles are typically considered though). These limitations may introduce potential inaccuracies in our results, particularly regarding low-energy albedo and secondary neutrons, although they are not expected to significantly affect the overall conclusions drawn from our study, such as material selection. It is acknowledged that neutrons significantly contribute to the radiation environment on Mars \cite{Clowdsley:2001}, and their interactions with shielding materials merit careful consideration. Utilizing specialized physics lists like the High Precision (HP) models, hybrid approaches combining Geant4 with neutron-specific codes like MCNPX, and validating simulations with experimental data can mitigate these issues. Future iterations of the model should aim to refine the simulation of neutron interactions. 

We now discuss other parameters that could affect the outcomes. Firstly, the Martian temperature at the equator can vary from -113$^{\circ}$ C during the night to 17$^{\circ}$ C during the day \cite{Barker:1998}. Additionally, the atmosphere experiences a thermal tide by the heat of the Sun on dayside and an infrared cooling on the nightside. According to \cite{Guo:2017} though, their effects in the radiation doses should be small compared to other factors like the solar modulation. Therefore, we do not investigate any variations of absorbed dose with respect to the temperature. 

It is also worth mentioning that the passing of cosmic rays through materials has an important effect on the nature of the materials. The most characteristic examples are the sensitivity of the electronic equipment to any kind of radiation, and of course, the harmful effect of rays to human organs, which rise to the creation of illnesses like tumors. An in-depth study of the energy distribution of charged particles on the Martian environment is maybe required to assess the damage on different substances, however, exploring these effects in our simulations is beyond the scopes of this paper.

The regolith's varying properties could potentially influence our decisions on where to locate a human habitat on Mars as well. For example, the value of its density has been derived in the literature from thermal inertia measurements, or indirectly from rover measurements or from assumptions about the grain density and the pore space, and there is a full range of values that can be attributed to regolith's density. The composition of Martian regolith can vary itself too; near the poles of the planet regolith is found in combination with water and ice, whereas near the equator it is mostly composed out of silicon. In other words, we cannot adopt one composition model, as the regolith's characteristics depend strongly on the location. Nevertheless, despite the fact the exact features of regolith may vary a little, it has been shown in previous studies about Martian regolith \cite{Kim:1998} and about Moon's regolith, which is comparable to Martian \cite{Nealy:1988,Nealy:1989} that small changes don’t affect significantly shielding properties. All subgroups of Martian soil have basically the same shield protection, and so, analyses for in-situ resources used for radiation protection can be reasonably made with a single Martian regolith model. This opinion is also supported by test simulations we performed with slightly different regolith compositions.

A practical consideration is the ease of implementation of different shielding materials. Adding a few centimeters of shielding material can be a critical factor in designing effective radiation protection strategies. For instance, our simulations add evidence that Martian regolith, with its inherent shielding properties, may offer a pragmatic solution due to its availability on the planet's surface. The relatively small differences in shielding effectiveness between materials, as indicated by our results, suggest that the choice of shielding material may be guided by factors such as ease of transport, construction, and adaptability to the Martian environment. The decision on shielding material should be a multifaceted one, taking into account not only the radiation attenuation capabilities but also the practical aspects of implementation.

In considering the practical implications of our simulations, the influence of the chosen shielding thickness of 10 g cm$^{-2}$ deserves discussion as well. While representing typical shielding requirements, thicker shielding, like 20 or 30 g/cm², could obscure reported effects, while thinner shielding might exaggerate material differences. A yet another reason behind the selected thickness of 10 g cm$^{-2}$ relies on a balance between realism and computational feasibility. Future model iterations should explore varying thicknesses to validate conclusions and enhance applicability in Martian scenarios, recognizing the inherent uncertainty introduced by the chosen thickness.

Lastly, we note that the discussion about the real effectiveness of passive shielding materials in space will become more established as soon as experiments will be compared against models. For instance, \cite{Sen:2010} performed experiments of polyethylene in a simulated Martian atmosphere and examined the fabrication of composite materials based on Martian regolith using polyethylene as a binding material. Inevitably, every coding model implicitly makes a number of necessary simplifications of the actual conditions of an environment, in order to keep simulation times manageable. Therefore, it is always possible that some of the theoretical conclusions could be premature or over-interpreted at some extent. Future work should shed more light on the field.

\section{Summary and Conclusions}
Protection from radiation is a top priority in planning long-duration missions in outer space. Especially on Mars, protecting astronauts from the hazardous cosmic rays is absolutely necessary due to the continual bombard of Galactic Cosmic Rays throughout the duration of the mission. A combination of measurements and model calculations in context of Mars are needed to fully understand the radiation effects on human health. Additionally, optimization and validation of radiation mitigation strategies is absolutely essential for preparing any human expedition of the red planet. 

There is an ongoing effort to find suitable materials to protect astronauts from radiation in deep space environments, like during the planned human journeys to Mars. Likewise, the MSL RAD measurements have allowed us to develop coding models of the Martian radiation environment and validate the obtained results. Nevertheless, there is a lack in literature in combining these two efforts, i.e. determining the behavior of passive shielding materials specifically in the Martian radiation field. This study aims to foster that need. We examined a number of materials in the Martian environment according to their passive shielding properties, using both GCR and some typical SPEs as input spectra. Apart from the most widely used materials in spacecrafts and missions so far, we also tested a range of innovative materials, not previously thoroughly analyzed for their radiation properties or considered as an option precisely on Mars. As a next step, we moved on to simulating combinations of materials with the common aluminum and the potentially useful Martian regolith.

For that purpose, we developed a model of Mars using the Geant4 toolkit, in which the planet was represented by a series of absorbing layers (atmosphere, shielding material, water as a substitute of a human, and soil). To gauge the accuracy of our model, we examined any sources of error and crosschecked the resulting dose rates with the respective values by RAD, finding a significant consistency. As a result,  although such a slab-plane model has a much simpler geometrical configuration compared to other published models used, we can safely argue that our model can nonetheless provide us with valuable answers to our research problem. Having an additional reliable and simple model of radiation on Mars, which can run under reasonable computational times, is quite valuable for a demanding and complicated task like protecting astronauts there from the dangerous cosmic radiation.

The optimal materials for space radiation shielding purposes should offer both effective stopping power and nuclear fragmentation. Our simulations suggest that the richer a material is in hydrogen, the more energy it absorbs. This is however only one of the parameters determining the efficiency of a substance against cosmic rays, another prominent one being the abundance in low atomic number elements. The most commonly used material in the aerospace field, aluminum, is not as effective as water, plastic, or compounds based on Li, but its still reduces a significant part of incoming radiation. We found that hydrogen and carbon-based compounds have an analogous behavior against cosmic rays, while materials made out of B or N are not as helpful. 
Lastly, we propose regolith combinations as an alternative option for shielding purposes, in order to keep launch mass requirements reasonable, although regolith's response against charged and energetic particles is not ideal, but sufficient enough. 

The in-depth study of the radiation environment on Mars is more than needed in order to plan and carry out successful human missions there. The engineering approach is to prepare all systems for the worst case scenario, therefore more work must be done in the space radiation field. Besides the obvious importance of this kind of this research for human spaceflight, we remark that studying the radiation field on Mars has many implications for astrobiology and the possible mechanisms through which life could possibly have emerged on the red planet.



\section*{Appendix}
There is not an absolute way to calculate radiation risk. Some physical quantities have been established to assess it. Usually, the following three ways are adopted to measure radiation:
\begin{itemize}
  \item Dose ($D$): It is defined as the energy absorbed per mass, regardless of the type of radiation. It is often measured in centi-Grays ($1\,cGy=0.01\,Gy=1\, rad$).
  \item Equivalent dose ($H$): It is represented by dose multiplied by a correction factor to account for the type of radiation. Its common unit is centi-Sieverts ($1\, cSv=0.01\, Sv=1\, rem$).
  \item Effective dose ($E$): It is defined as equivalent dose multiplied by a factor considering the part of the body irradiated. 
\end{itemize}

More precisely, for all $i$ particles, each one of energy $E$, the dose at depth $x$ is given by:

\begin{equation} 
    D=\sum_{i} \int_{0}^{\infty} S_i(E) \Phi_i(x, E)dE 
\end{equation}

$S_i(E)$ is the stopping power of the material and $\Phi_i(x, E)$ is the flux of particles. Similarly, the equivalent dose is calculated by multiplying the above relation with the weighting factor $Q_i(E)$ corresponding to the particle type, as mentioned previously:

\begin{equation} 
    H=\sum_{i} \int_{0}^{\infty} Q_i(E) S_i(E) \Phi_i(x, E)dE 
\end{equation}

$Q_i(E)$ is computed by the linear energy transfer of each particle. Lastly, the total effective dose for a human body is the weighted average for all organs (subscript $o$ stands for organ and $w$ is the weighting factor \cite{ICRP:2013}):

\begin{equation} 
    E=\sum_{o} w_o H_o 
\end{equation}

Hence, radiation quantities are corrected according to factors like the energy spectrum intensity, composition and the tissue(s) impacted. Still, many uncertainties arise when determining the radiation effects on astronauts. Radiation risk varies for example as per other aspects like age, gender or individual susceptibility.

The quality factor Q is introduced due to the fact that different radiation types are not equally hazardous, even if they deposit the same energy. For the dose equivalent, x-rays stand as the reference radiation type, at $Q=1$. Electrons and photons have the same $Q$, while for protons it is dependent on their LET. On the other hand, helium nuclei are much more dangerous ($Q=20$), which is also the value for 1 MeV neutrons. Neutrons' $Q$ is decreasing for other energies. In general, the absorbed dose is related to the energy lost by the incident particles, although there are some exceptions to this behavior.

Finally, ICRP Publication 60 \cite{ICRP:1990} introduced a mean radiation quality factor $<Q>$ to distinguish between different radiation environments and gauge the severity of them. $<Q>$ is calculated by dividing the total dose equivalent by the total absorbed dose.

\section*{Acknowledgements}

The authors are grateful to two anonymous referees for their constructive comments that improved this work. We thank the members of the Geant4 collaboration who developed the software (geant4.cern.ch) used for this work. This work was supported by the New York University Abu Dhabi (NYUAD) Institute Research Grants G1502 and CG014 and the ASPIRE Award for Research Excellence (AARE) Grant S1560 by the Advanced Technology Research Council (ATRC).

\section*{Author contribution statement}

Conceptualization, D.G., D.A.; writing—original draft preparation, D.G., D.A.; writing—review and editing, D.G., D.A.. All authors reviewed the manuscript. 

\section*{Competing interests}
The authors declare that they have no conflict of interest.

\section*{Data availability statement}
The codes and the data that were used to prepare our models within the paper are available from the corresponding authors upon reasonable request.








\bibliography{sample}

\end{document}